\documentclass[12pt,draftclsnofoot, onecolumn]{IEEEtran}

\usepackage{graphicx}
\usepackage{amsmath}
\usepackage{amssymb}
\usepackage[caption=false]{subfig}
\usepackage[noadjust]{cite}
\usepackage{float}
\usepackage{algorithm}
\usepackage{bibentry}
\usepackage{balance}
\usepackage{algorithm}
\usepackage{algorithmicx}
\usepackage{algpseudocode}
\usepackage{xcolor}
\usepackage{balance}
\usepackage{multirow}

\newtheorem{theorem}{Theorem}
\newtheorem{remark}{Remark}

\DeclareMathOperator{\Tr}{\rm tr}

\graphicspath{{img/}}

\begin{document}

\title{Max-Min Fair Beamforming for Cooperative Multigroup Multicasting with Rate-Splitting}

\author{Ahmet Zahid Yal\c{c}{\i}n and Yavuz Yap{\i}c{\i}}

\maketitle

\begin{abstract}
The demand of massive access to the same multimedia content at the same time is one major challenge for next-generation cellular networks in densely-packed urban areas. The content-aware multicast transmission strategies provide promising solutions to such use cases involving many mobile users trying to fetch the same data. In this work, we consider multigroup multicast transmission with a common message (e.g., multimedia content), in which different multicast groups are interested along with their private multicast messages. We further assume that a relay helps the cellular base station (BS) disseminate multicast content to the users experiencing high path loss and/or blockage. We propose superposition and concatenated coding schemes, denoted by SC and CC, respectively, to transmit the common and private multicast messages. In order to maximize max-min fair (MMF) rates, we design a novel low-complexity alternating-optimization algorithm to compute transmit and relay precoders. We also propose rate-splitting (RS) alternatives of SC and CC schemes together with an iterative algorithm to derive dedicated transmit and relay precoders so as to maximize MMF rates. We rigorously evaluate the performance of the proposed transmission schemes and precoders, and verify the superiority of RS-based schemes in overloaded scenarios without any saturation with increasing signal-to-noise ratio.
\end{abstract}


\section{Introduction}
\label{sec:introduction}

The mobile data traffic has risen unprecedentedly over the past several years, with a growth rate of $78\%$ just between the second quarters of 2018 and 2019 \cite{ericsson2019mobrep}. This huge amount of mobile traffic is currently well above 32 exabytes (EB), and is forecast to exceed 930 EB by 2022 \cite{Cisco2019visnet}. The main driving source of this enormous mobile data traffic comes from multimedia contents shared over the cellular network (e.g., streaming or broadcasting live videos on social media). In such use cases, multiple---possibly many---users dynamically request to access to the same mobile content simultaneously. One promising solution to handle such a massive access demand is to develop multicast strategies which enable transmission of a common message (e.g., multimedia content) to many users simultaneously \cite{Sidiropoulos2006_TransmitBeamforming, multicast_2, Jindal2006_CapacityLimitsofMultipleAnt, Abdelkader2010_MultipleAntennaMulticasting, Chiu2009_TransmitPrecoding}. Assuming heterogeneity of mobile users with diverse needs, group-oriented service designs are yet another appealing extension to this point-to-multipoint transmission strategy where multiple multicast messages are sent to different user groups at the same time, which is referred to as multigroup multicasting~\cite{Silva2009_LinearTransmitBeamforming,Qiu2018ResAll,Araniti2018MulMul}. Content-oriented mobile services (e.g., application updates, wireless caching) can be optimized by this way assuming similar contents are required by the users of the same group \cite{Karidipis2008_QoS,Araniti2017}. 

Along with the use of multiple antennas at the cellular base stations (BSs), optimal precoding design for multigroup multicasting becomes a challenging task since the inherent inter-group interference should be handled carefully. There are various studies in the literature which consider multigroup multicasting precoding with different optimization goals involving maximization of weighted sum-rate (WSR), max-min fairness (MMF), and minimization of transmit power. In particular, \cite{Karidipis2008_QoSMaxMinFairTransmit} investigates precoding strategies that aims at guaranteed quality of service (QoS) with minimum transmit power budget while \cite{Christopoulos2014_WeightedFair} proceeds to solve the precoding problem under per antenna power constraints. In \cite{Schad2012}, the problem of maximizing the minimum signal-to-interference-plus-noise ratio (SINR) of all the receivers is investigated under power constraints, and a sequential convex programming algorithm is developed to obtain an MMF-optimal precoder. 

A content-centric transmission scheme is considered in \cite{Tao2016_ContentCentric} for a cloud radio access network (C-RAN), which proposes a transmit beamforming design based on multigroup multicasting and caching. The maximization of minimum group rate is investigated in \cite{Joudeh2017_RateSplittingforMaxMin} for multigroup multicast transmission with rate splitting (RS), which is employed to combat inter-group interference. \cite{Venkatraman2017} considers the precoder design problem for multigroup multicasting in a multi-input multi-output (MIMO) orthogonal frequency-division multiplexing (OFDM) framework, which aims at either minimizing the total transmit power under QoS constraints, or maximizing the minimum rate among users by successive convex approximation (SCA). In \cite{Yalcin2019_MGMC_withCommon}, a common message is assumed across the multicast groups, and the optimal precoder that maximizes WSR is computed for this overlapping (i.e., due to the common message) multigroup multicast scenario.

In an effort to disseminate multicast content to distant mobile users experiencing high path loss (PL), or those without strong line-of-sight (LoS) links, relay-aided multicasting schemes have recently attracted much attention as an appealing research direction. The problem of optimal precoder design for relay-aided multicasting schemes is even more challenging (in comparison to no-relay schemes) as it requires joint optimization of transmit and relay precoders, each of which leads to non-convex objective functions. In \cite{Bornhorst2012}, multigroup multicasting is considered for a relay-aided multiple transmitter and relay scenario, and a distributed beamforming algorithm is proposed which minimizes the total relay power. 

A cognitive cooperative multicast transmission scheme is considered in \cite{Lin2015DisCo} with the objective of maximizing the aggregate transmission rate by optimizing the relay precoders. In \cite{Ma2016StocBeam}, the authors consider an MMF-based precoder design for a two-hop amplify-and-forward (AF) relay network for multigroup multicast transmission. A multigroup multi-way relaying network is considered in \cite{Klein2017MulInt}, where the beamforming matrices are designed based on singular value decomposition (SVD) assuming perfect availability of channel state information (CSI). In \cite{Zhu2018}, the authors investigate the downlink cooperative multigroup multicast transmission in a hybrid terrestrial-satellite network, where the satellite (as transmitter) and base stations (as relay) provide the multicast service for ground users in a cooperative manner. 

% \cite{Rong2011_SimplifiedAlgo} proposes an algorithm to minimize the mean squared error (MSE) under the transmit (source) and relay power constraints. The problem of precoder design for a two-way relay-aided multi-input multi-output (MIMO) system is considered in \cite{He2014_JointTranceiver}, which aims at minimizing the total transmission power under MSE constraints. In \cite{Khandaker2013_PrecodingDesign, Khandaker2014}, transmitter and relay precoder design problem is investigated for multicasting over MIMO transmission based on amplify-and-forward (AF) relaying, where the optimization aims at minimizing the maximum MSE. A robust transceiver design is proposed by \cite{Gopal2017_RobustMMSE} for the system model of \cite{Khandaker2013_PrecodingDesign} to combat the mismatch in the channel state information (CSI).

% Max-Min Fair Beamforming for Cooperative Multigroup Multicasting with Rate-Splitting

In this work, we consider beamforming design for multigroup multicast transmission in a cooperative (i.e., relay-aided) MIMO network. In particular, we assume that users of each multicast group request not only a private message (i.e., multimedia content) specific to that group only, but also a \textit{common message} that is being asked for by the other multicast groups (i.e., overlapping multicast groups), as well. To the best of authors' knowledge, this multigroup multicast scenario of practical importance for future wireless networks is not considered elsewhere before. Our specific contributions are summarized in the following.
\begin{itemize}

    \item[---] We propose two frame structures to transmit the common message (to all the multicast groups): superposition coding (SC) and concatenated coding (CC). While the BS transmits the superposition of the common and private multicast messages in SC scheme, common message is treated as another private multicast message in CC scheme. We numerically show that CC is superior to SC in terms of both the convergence speed and MMF rates when it is combined with RS approach.
    
    \item[---] We propose an alternating-optimization algorithm to jointly design transmit and relay precoders so as to maximize the MMF rates for both SC and CC schemes. Since the MMF-rate maximization is a challenging non-convex problem, we propose an equivalent MMF optimization aiming at minimizing weighted mean squared error (WMSE) as a low-complexity solution. We theoretically prove the equivalency of these two approaches (at the optimal point), and verify the promised performance through numerical results.
    
    \item[---] We also propose two RS schemes (i.e., RS-SC and RS-CC) based on the frame structures of SC and CC. We derive an alternating-optimization algorithm for these RS schemes, which is based on MMF WMSE minimization. We numerically show that RS-based schemes significantly outperform non-RS schemes (i.e., SC and CC), especially in the \textit{overloaded} scenarios (with relatively more users). In addition, we show that non-RS schemes saturate for overloaded scenarios after a modest transmit SNR while RS-based schemes exhibit a linear increase. We also investigate the impact of the relay location for all these schemes.
    
\end{itemize}

% --- Note that when there is no common message to broadcast, RS-based schemes achieve an MMF rate that is roughly twice as large as that of non-RS schemes. 
% --- RS promises a superior performance to NOMA as RS proceeds to decode the multiuser interference partially (as opposed to NOMA trying to decode all the interference for the strong user), and hence is expected to manifest its strength under high multiuser interference scenarios.
% --- Alternating-optimization based iterative solution, which converges very quickly.  
%  and is concatenated to the multicast message stream.
% --- CC is superior to SC: The converge rate and MMF-rate performance of RS-CC is better than that of RS-SC. (though there is almost no difference between the SC and CC schemes without RS)

{\color{black}Note that the 3rd Generation Partnership Project (3GPP) has recently started working on Release-17 of the new radio (NR) specification (for 5G and beyond) that will include the \textit{physical layer multicasting} support \cite{3GPP_Rel17}, which is pointing out the \textit{significance} and \textit{timeliness} of our study for communications literature.} The rest of the paper is organized as follows. Section~\ref{sec:system_model} introduces the system and signal model. Section~\ref{sec:wmmse} considers equivalent MMF WMSE problem, and the iterative precoder design is given in Section~\ref{sec:precoder_design}. The RS-based transmission schemes and precoder design are described in Section~\ref{sec:rate_splitting}, and the numerical results are presented in Section~\ref{sec:results}. The paper concludes with some final remarks in Section~\ref{sec:conclusion}.

\textit{Notation:} $\|\cdot\|_\mathsf{F}$, $\bigcup$, $\bigcap$, $\mathbb{E}$, and $\emptyset$ stand for the Frobenius norm, union of sets, intersection of sets, statistical expectation, and the empty set, respectively. $|\cdot|$ is the cardinality of a set or norm of a complex-valued quantity. $\mathbb{C}^{M{\times}N}$ denotes complex-valued matrices of size ${M{\times}N}$.

\section{System Model }\label{sec:system_model}

\begin{figure}[!t]
\centering
\includegraphics[width=0.65\textwidth]{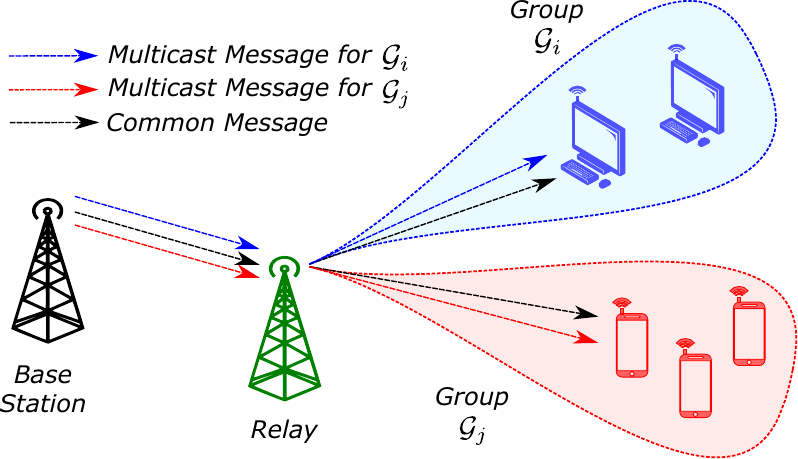}
\caption{System model for relay-aided multigroup multicast scenario with the groups $i$ and $j$ (i.e., $\mathcal{G}_i$ and $\mathcal{G}_j$).} \label{fig:system_model}
\centering
\end{figure}

We consider a multi-antenna relay network which comprises of a single BS equipped with $M$ antennas, $N$ single-antenna receivers indexed by the set $\mathcal{N} \,{=}\,  \{1, \dots , N\}$, and a multi-antenna relay equipped with $N_\mathsf{R}$ antennas{\color{black}, as shown in Fig.~\ref{fig:system_model}}. The receivers are grouped into the $K$ multicast groups $\mathcal{G}_1,\ldots,\mathcal{G}_K$ with $1 \,{\leq}\, K \,{\leq}\, N$, where $\mathcal{G}_k$ is the set of receivers belonging to the $k$-th group with $k \, {\in} \, \mathcal{K} \,{=}\, \{1,\dots,K\}$. We assume that each receiver belongs to exactly one group so that $\bigcup_{k\in \mathcal{K}} \mathcal{G}_k \,{=}\, \mathcal{N}$ and $\mathcal{G}_k \bigcap \mathcal{G}_j \,{=}\, \emptyset$, $\forall k,j \,{\in}\, \mathcal{K}$ and $k \neq j$. In addition, we define the inverse function $\mu \,{:}\, \mathcal{N} \,{\rightarrow}\, \mathcal{K}$ to map the users to their respective multicast groups such that $\mu(n) \,{=}\, k$ for all $n \,{\in}\, \mathcal{G}_k$. Denoting the number of users in the $k$-th group by $G_k \,{=}\, |\mathcal{G}_k|$, we assume without any loss of generality that the group sizes are in an ascending order such that
\begin{align}
G_1 \leq G_2 \leq \dots \leq G_K .
\end{align}

\subsection{Transmit and Receive Signal Models}
\label{sec:system_model_signal}

We assume that the BS transmits a common message intended for all $N$ users, and $K$ multicast messages each of which targets each of the $K$ multicast groups. The single common and multiple multicast messages are encoded into the streams $s_\mathsf{c}$ and $s_{\mathsf{u},1} \dots s_{\mathsf{u},K}$, respectively. The overall transmission is accomplished by employing either superposition coding (SC) or concatenated coding (CC) schemes at the BS, which are detailed in the following.

The SC scheme generates a transmit message vector $\textbf{s}^\mathsf{SC} \,{=}\, \left[s_1 \dots s_K\right]^{\rm T} {\in}\, \mathbb{C}^{K{\times}1}$, which consists of superimposed common and multicast messages such that $\textbf{s}^\mathsf{SC} \,{=}\, $ $\textbf{s}^\mathsf{SC}_\mathsf{c} \,{+}\, \textbf{s}^\mathsf{SC}_\mathsf{u}$ where $\textbf{s}^\mathsf{SC}_\mathsf{c} \,{=}\, \left[s_\mathsf{c} \dots s_\mathsf{c}\right]^{\rm T}$ and $\textbf{s}^\mathsf{SC}_\mathsf{u} \,{=}\, \left[ s_{\mathsf{u},1} \dots s_{\mathsf{u},K}\right]^{\rm T}$. We assume that $\mathbb{E}\{|s_c|^2\} \,{=}\, \alpha$ and $\mathbb{E}\{s_{\mathsf{u},k} s_{\mathsf{u},\ell}^*\} \,{=}\,\bar{\alpha} \, \delta_{k\ell}$ with $\bar{\alpha} \,{=}\, 1 \,{-}\, \alpha$, where $\alpha$ is the ratio of power allocated to common message. The transmit message vector is then linearly processed by a precoder matrix $\textbf{F}^\mathsf{SC} \,{=}\, \left[ \textbf{f}_1^\mathsf{\,SC} \dots \textbf{f}_{K}^\mathsf{\,SC} \right]$  $\in \mathbb{C}^{M{\times}K}$, where $\textbf{f}_k^\mathsf{\,SC}$ is the precoder vector for the superimposed message for the $k$-th multicast group. 

The CC scheme, on the other hand, treats common message similar to each multicast message such that a dedicated precoder is considered for the transmission of common message. More specifically, the transmit message vector is given by $\textbf{s}^\mathsf{CC} \,{=}\, \left[s_\mathsf{c} \, s_{\mathsf{u},1} \dots s_{\mathsf{u},K} \right]^{\rm T} {\in}\, \mathbb{C}^{(K{+}1){\times}1}$ such that  $\mathbb{E}\{|s_c|^2\} \,{=}\, 1$ and $\mathbb{E}\{s_{\mathsf{u},k} s_{\mathsf{u},\ell}^*\} \,{=}\, \delta_{k\ell}$. The respective linear precoder is then given as $\textbf{F}^\mathsf{CC} \,{=}\, \left[ \textbf{f}_\mathsf{c}^\mathsf{\,CC} \, \textbf{f}_1^\mathsf{\,CC} \dots \textbf{f}_{K}^\mathsf{\,CC} \right]$  $\in \mathbb{C}^{M{\times}(K{+}1)}$, where $\textbf{f}_\mathsf{c}^\mathsf{\,CC}$ and $\textbf{f}_k^\mathsf{\,CC}$ are the precoder vectors for the common and multicast message of the $k$-th group, respectively. 

In the first time slot of the overall transmission mechanism, the BS transmits the message vector $\textbf{x}_\mathsf{T}^\mathsf{t} \,{\in}\, \mathbb{C}^{M{\times}1}$ with $\mathsf{t} \,{\in}\, \{\mathsf{SC,CC}\}$ to the relay terminal, which can be given compactly as
\begin{align}
\textbf{x}_\mathsf{T}^\mathsf{t} &= \textbf{F}^\mathsf{t} \textbf{s}^\mathsf{t} = \textbf{f}_\mathsf{c}^\mathsf{\,t} s_\mathsf{c} + \sum_{k \, {\in} \, \mathcal{K}} \textbf{f}_k^\mathsf{\,t} s_{\mathsf{u},k}, \label{precodedSignal}
\end{align}
where $\textbf{f}_\mathsf{c}^\mathsf{\,SC} \,{=}\, \sum_{k=1}^{K}\textbf{f}_k^\mathsf{SC}$. The respective average power constraint at the BS is given as
\begin{align} \label{pow_const_tx}
    \mathsf{B}^\mathsf{t} \|\textbf{f}_c^\mathsf{\,t} \|^2_\mathsf{F} + \mathsf{C}^\mathsf{t} \sum_{k \, {\in} \, \mathcal{K}} \|\textbf{f}_k^\mathsf{\,t} \|^2_\mathsf{F} \leq \mathsf{P}_\mathsf{tx}, 
\end{align}
where $\mathsf{P}_\mathsf{tx}$ denotes the total transmit power at the BS, with the coefficients $(\mathsf{B}^\mathsf{sc} ,\mathsf{C}^\mathsf{sc}) \,{=}\, (\alpha, \bar{\alpha})$ and $(\mathsf{B}^\mathsf{cc} ,\mathsf{C}^\mathsf{cc}) \,{=}\, (1,1)$. The received signal at the relay terminal is then given by
\begin{align}\label{rec_signal_1}
 \textbf{y}_\mathsf{R}^\mathsf{t} &= \textbf{H}_\mathsf{SR} \, \textbf{x}_\mathsf{T}^\mathsf{t} + \textbf{n}_\mathsf{R} = \textbf{H}_\mathsf{SR} \textbf{f}_\mathsf{c}^\mathsf{\,t} s_\mathsf{c} + \textbf{H}_\mathsf{SR} \sum_{k \, {\in} \, \mathcal{K}} \textbf{f}_k^\mathsf{\,t}  s_{\mathsf{u},k} + \textbf{n}_\mathsf{R}, 
\end{align}
where ${\textbf{H}_\mathsf{SR}} \,{\in}\, \mathbb{C}^{N_\mathsf{R}{\times}M}$ is the channel between the BS (source) and the relay, which is composed of independent standard complex Gaussian entries, and $\textbf{n}_\mathsf{R}$ is the observation noise being zero-mean Gaussian with the covariance $\sigma^2\textbf{I}$. In addition, we define the transmit signal-to-noise ratio (SNR) as $\gamma \,{=}\, \mathsf{P}_\mathsf{tx}/\sigma^2$.

In the second time slot, the relay forwards the received signal $\textbf{y}_\mathsf{R}$ to the users (destination) by the relay precoder $\textbf{G}^\mathsf{t} \,{\in}\, \mathbb{C}^{N_\mathsf{R} {\times} N_\mathsf{R}}$ based on AF strategy. The respective message vector $\textbf{x}_\mathsf{R}^\mathsf{t}$ to be transmitted by the relay is then given as
\begin{align}
\textbf{x}_\mathsf{R}^\mathsf{t} &=  \textbf{G}^\mathsf{t}  \textbf{H}_\mathsf{SR} \textbf{f}_\mathsf{c}^\mathsf{\,t} s_\mathsf{c} + \textbf{G}^\mathsf{t} \textbf{H}_\mathsf{SR} \sum_{k \, {\in} \, \mathcal{K}} \textbf{f}_k^\mathsf{\,t}  s_{\mathsf{u},k} + \textbf{G}^\mathsf{t} \textbf{n}_\mathsf{R} ,
\end{align} 
with the average power constraint at the relay 
\begin{align}
    \mathsf{B}^\mathsf{t} \|\textbf{G}^\mathsf{t} \textbf{H}_\mathsf{SR} \textbf{f}_\mathsf{c}^\mathsf{\,t} \|^2_\mathsf{F} 
    + \mathsf{C}^\mathsf{t} \sum_{k \, {\in} \, \mathcal{K}} \| \textbf{G}^\mathsf{t} \textbf{H}_\mathsf{SR} \textbf{f}_k^\mathsf{\,t} \|^2_\mathsf{F} 
    + \sigma^2 \|\textbf{G}^\mathsf{t} \|^2_\mathsf{F} \leq \mathsf{P}_\mathsf{re}, \label{pow_const_rel}
\end{align}
where $\mathsf{P}_\mathsf{re}$ denotes the total transmit power at the relay. The received signal at the $n$-th user is
\begin{align}
 y_{\mathsf{U},n}^\mathsf{t} &= \textbf{h}_n \textbf{G}^\mathsf{t} \textbf{H}_\mathsf{SR} \textbf{f}_\mathsf{c}^\mathsf{\,t} s_\mathsf{c} + \textbf{h}_n \textbf{G}^\mathsf{t} \textbf{H}_\mathsf{SR} \sum_{k \, {\in} \, \mathcal{K}} \textbf{f}_k^\mathsf{\,t}  s_{\mathsf{u},k} + \textbf{h}_n \textbf{G}^\mathsf{t} \textbf{n}_\mathsf{R} + n_{\mathsf{U},n},  \label{rec_signal_2}
\end{align}
where $\textbf{h}_n \,{\in}\, \mathbb{C}^{1{\times}N_\mathsf{R}}$ is the channel between the relay and the $n$-th user, and $n_{\mathsf{U},n}$ is the respective observation noise. Without any loss of generality, we assume that the entries of $\textbf{h}_n$ and $n_{\mathsf{U},n}$ are all independent complex Gaussian with zero mean and unit variance. 

\subsection{Achievable Rates and Problem Definition}

In the proposed transmission scheme, each user should first decode the common message, and then subtract it from its received signal as per successive interference cancellation (SIC) strategy. The multicast messages are then decoded by the respective users. The achievable rates associated with common and multicast messages are given, respectively, for the $n$-th user as %follows
\begin{align}
\mathsf{R}_{\mathsf{c},n}^\mathsf{t} &= \log \left( 1 + \mathsf{B}^\mathsf{t} \sigma^{{-}2}_{\mathsf{c},n} \left|\textbf{h}_n \textbf{G}^\mathsf{t} \textbf{H}_\mathsf{SR} \textbf{f}_\mathsf{c}^\mathsf{\,t} \right|^2  \right),\label{eq:achievable_rate_common}\\
\mathsf{R}_{\mathsf{u},n}^\mathsf{t} &= \log \left( {1} + \mathsf{C}^\mathsf{t} \sigma^{{-}2}_{\mathsf{u},n} \left|\textbf{h}_n \textbf{G}^\mathsf{t} \textbf{H}_\mathsf{SR} \textbf{f}_{\mu(n)}^\mathsf{\,t} \right|^2  \right)\label{eq:achievable_rate_unicast},
\end{align}
where $\sigma^2_{\mathsf{c},n}$ and $\sigma^2_{\mathsf{u},n}$ are the effective noise variances of common and multicast messages for the $n$-th user, which are given as
\begin{align}
\sigma^2_{\mathsf{c},n} &= \sum_{i \,{\in}\, \mathcal{K}} \mathsf{C}^\mathsf{t} \left|\textbf{h}_n \textbf{G}^\mathsf{t} \textbf{H}_\mathsf{SR} \textbf{f}_i^\mathsf{\,t}\right|^2 
+ \sigma^2 \left( 1 + \left|\textbf{h}_n \textbf{G}^\mathsf{t}\right|^2 \right),\\
\sigma^2_{\mathsf{u},n} &= \sigma^2_{\mathsf{c},n} - \mathsf{C}^\mathsf{t} \left|\textbf{h}_n \textbf{G}^\mathsf{t} \textbf{H}_\mathsf{SR} \textbf{f}_{\mu(n)}^\mathsf{\,t}\right|^2.
\end{align}
Since the common message should be decoded by each user, and the $k$-th multicast message should be decoded by each user in only the $k$-th multicast group, the overall achievable rate associated with the common and multicast messages are given, respectively, as follows
\begin{align} 
\mathsf{R}_\mathsf{C}^\mathsf{t} &= \min_{n \,{\in}\, \mathcal{N}} \mathsf{R}_{\mathsf{c},n}^\mathsf{t},\label{eqn:minRc} \\
\mathsf{R}_k^\mathsf{t} &= \min_{n \,{\in}\, \mathcal{G}_k} \mathsf{R}_{\mathsf{u},n}^\mathsf{t} . \label{eqn:minRu}
\end{align} 

Our objective is to obtain the optimal transmit and relay precoders which maximize the minimum multicast group rate. In other words, the goal of the underlying precoder design is to achieve fairness among multicast groups subject to the transmit and  the relay power constraints. We therefore formulate the respective optimization problem as a linear MMF design, which is
\begin{IEEEeqnarray}{rl}
\max_{\textbf{F}^\mathsf{t}, \textbf{G}^\mathsf{t}}
&\qquad \min_{k \,{\in}\, \mathcal{K}} \min_{n \,{\in}\, \mathcal{G}_k} \mathsf{R}_{\mathsf{u},n}^\mathsf{t} \label{eq:optimization_MMFSR1} \\
\text{s.t.}& \qquad \mathsf{R}_\mathsf{c}^\mathsf{th} \leq \mathsf{R}_{\mathsf{c},n}^\mathsf{t}   , \, \forall n \,{\in}\, \mathcal{N} , \IEEEyessubnumber\\
&\qquad \eqref{pow_const_tx},  \eqref{pow_const_rel},
\IEEEyessubnumber \label{eq:optimization_MMFWSR1_power}
\end{IEEEeqnarray}
where $\mathsf{R}_\mathsf{c}^\mathsf{th}$ stands for the threshold rate for the common message constituting the common rate constraint, and \eqref{eq:optimization_MMFWSR1_power} represents the power constraints at the transmitter and relay ends. Note that the optimization problem in \eqref{eq:optimization_MMFSR1} is non-convex and difficult to solve. In an attempt to obtain a solution, we convert this optimization to a smooth constrained optimization problem by defining a new auxiliary variable $\theta$, which is given as follows
\begin{IEEEeqnarray}{rl}
\max_{\textbf{F}^\mathsf{t}, \textbf{G}^\mathsf{t},\theta}
&\qquad \theta \label{eq:optimization_MMFSR2} \\
\text{s.t.} & \qquad \theta \leq \mathsf{R}_{\mathsf{u},n}^\mathsf{t}, \, {\color{black} \forall n} \,{\in}\, \mathcal{N}, \IEEEyessubnumber \label{eq:optimization_MMFWSR2_multicast}\\
& \qquad \mathsf{R}_\mathsf{c}^\mathsf{th} \leq \mathsf{R}_{\mathsf{c},n}^\mathsf{t}   , \, \forall n \,{\in}\, \mathcal{N} , \IEEEyessubnumber\\
&\qquad \eqref{pow_const_tx},  \eqref{pow_const_rel},
\IEEEyessubnumber \label{eq:optimization_MMFWSR2_power}
\end{IEEEeqnarray}
In the following, we consider an alternative strategy to obtain a solution for the MMF problem \eqref{eq:optimization_MMFSR1} and its modified version \eqref{eq:optimization_MMFSR2}.

\section{Equivalent MMF WMSE Problem}
\label{sec:wmmse}

In this section, we derive the equivalent WMSE expressions for the communications scenario in hand, which are used in the next section to develop an iterative precoder design based on the relation between the mutual information (i.e., achievable rates) and WMSE \cite{Paulraj2001GenLin,guo2005mutual,Utschick2006AltOpt,Christensen2008_WeightedSumRate,Luo2011WMMSE}. We adopt the same message decoding strategy of Section~\ref{sec:system_model}. In particular, the weighted minimum MSE (WMMSE) receiver first processes the received signal $y_{\mathsf{U},n}^\mathsf{t}$ of the $n$-th user by $W_n^\mathsf{t}$ to obtain an estimate of the common message, which is given as $\hat{s}_{\mathsf{c},n}^\mathsf{t} \,{=}\, W_n^\mathsf{t} \, y_{\mathsf{U},n}^\mathsf{t}$. Once the estimate of the common message is cancelled from the received signal assuming perfect SIC, the estimate of the multicast message is computed as $\hat{s}_{\mathsf{u},n}^\mathsf{t} \,{=}\, V_n^\mathsf{t} \left(y_{\mathsf{U},n}^\mathsf{t} \,{-}\, \textbf{h}_n \textbf{G}^\mathsf{t} \textbf{H}_\mathsf{SR} \textbf{f}_\mathsf{c}^\mathsf{\,t} s_\mathsf{c} \right)$. The mean square error (MSE) of estimating the common and multicast messages are defined as $\varepsilon_{\mathsf{c},n}^\mathsf{t} \,{=}\, \mathbb{E}\left\lbrace \left|\hat{s}_{\mathsf{c},n}^\mathsf{t} \,{-}\, s_c\right|^2\right\rbrace$ and $\varepsilon_{\mathsf{u},n}^\mathsf{t} \,{=}\, \mathbb{E}\left\{\left|\hat s_{\mathsf{u},n}^\mathsf{t} \,{-}\, s_{\mathsf{u},n} \right|^2\right\}$, respectively, which can readily be represented as
\begin{align}
\varepsilon_{\mathsf{c},n}^\mathsf{t} &= \left| W_n^\mathsf{t} \right|^2 \left( \!\mathsf{C}^\mathsf{t} \!\sum_{i \in \mathcal{K}}  \|\textbf{h}_n \textbf{G}^\mathsf{t}  \textbf{H}_\mathsf{SR} \textbf{f}_i^\mathsf{t} \|^2_\mathsf{F} + \mathsf{B}^\mathsf{t} \| \textbf{h}_n \textbf{G}^\mathsf{t} \textbf{H}_\mathsf{SR} \textbf{f}_\mathsf{c}^\mathsf{\,t} \|^2_\mathsf{F} + \sigma^2 \| \textbf{h}_n \textbf{G}^\mathsf{t} \|^2_\mathsf{F} + \sigma^2 \right) \nonumber \\
&\quad + \mathsf{B}^\mathsf{t} \left( 1 - 2 \Tr\!\left( \mathcal{R} \left\lbrace W_{n}^\mathsf{t} \textbf{h}_n \textbf{G}^\mathsf{t} \textbf{H}_\mathsf{SR} \textbf{f}_\mathsf{c}^\mathsf{\,t} \right\rbrace \right) \right) ,  \label{eq:mse_common}\\
\varepsilon_{\mathsf{u},n}^\mathsf{t} &= \left| V_n^\mathsf{t} \right|^2 \left( \! \mathsf{C}^\mathsf{t} \! \sum_{i \in \mathcal{K}}  \|\textbf{h}_n \textbf{G}^\mathsf{t}  \textbf{H}_\mathsf{SR} \textbf{f}_i^\mathsf{t} \|^2_\mathsf{F} + \sigma^2 \|\textbf{h}_n \textbf{G}^\mathsf{t} \|^2_\mathsf{F}  + \sigma^2 \right) \nonumber \\
& \quad + \mathsf{C}^\mathsf{t} \left( 1 
\,{-}\,  2 \Tr\!\left( \mathcal{R} \left\lbrace V_n^\mathsf{t} \textbf{h}_n \textbf{G}^\mathsf{t}  \textbf{H}_\mathsf{SR} \textbf{f}_{\mu(n)}^\mathsf{\,t}  \right\rbrace \right)
\right). \label{eq:mse_unicast}
\end{align}
The optimal minimum MSE (MMSE) receivers for common and multicast messages are defined accordingly as the ones minimizing \eqref{eq:mse_common} and \eqref{eq:mse_unicast}, which are readily given, respectively, as \cite{Viswanath1999OptSeq}
\begin{align}
W_n^\mathsf{t,opt} &=  \mathsf{B}^\mathsf{t} \left( \textbf{h}_n \textbf{G}^\mathsf{t} \textbf{H}_\mathsf{SR} \textbf{f}_\mathsf{c}^\mathsf{\,t} \right)^{\rm H} \left( \mathsf{B}^\mathsf{t} \left| \textbf{h}_n \textbf{G}^\mathsf{t} \textbf{H}_\mathsf{SR} \textbf{f}_\mathsf{c}^\mathsf{\,t} \right|^2 + \sigma^2_{\mathsf{c},n} \right)^{-1},\label{eq:mmse_receiver_common}\\
V_n^\mathsf{t,opt} &= \mathsf{C}^\mathsf{t} \left( \textbf{h}_n \textbf{G}^\mathsf{t} \textbf{H}_\mathsf{SR} \textbf{f}_{\mu(n)}^\mathsf{\,t} \right)^{\rm H}  \left( \mathsf{C}^\mathsf{t} \left| \textbf{h}_n \textbf{G}^\mathsf{t} \textbf{H}_\mathsf{SR} \textbf{f}_{\mu(n)}^\mathsf{\,t} \right|^2 + \sigma^2_{\mathsf{u},n} \right)^{-1}, \label{eq:mmse_receiver_unicast}
\end{align}
and the respective MMSE values (i.e., \eqref{eq:mse_common} and \eqref{eq:mse_unicast} after employing \eqref{eq:mmse_receiver_common} and \eqref{eq:mmse_receiver_unicast}, respectively) are found to be
\begin{align}
\varepsilon_{\mathsf{c},n}^\mathsf{t,min} &=\left( {\frac{1}{\mathsf{B}^\mathsf{t}}} + \frac{1}{\sigma^2_{\mathsf{c},n}} \left|\textbf{h}_n \textbf{G}^\mathsf{t} \textbf{H}_\mathsf{SR} \textbf{f}_\mathsf{c}^\mathsf{\,t}\right|^2 \right)^{{-}1}, \label{eq:mmse_common}\\
\varepsilon_{\mathsf{u},n}^\mathsf{t,min} &= \left( {\frac{1}{\mathsf{C}^\mathsf{t}}} + \frac{1}{\sigma^2_{\mathsf{u},n}} \left|\textbf{h}_n \textbf{G}^\mathsf{t} \textbf{H}_\mathsf{SR} \textbf{f}_{\mu(n)}^\mathsf{\,t}\right|^2 \right)^{-1}.\label{eq:mmse_unicast}
\end{align}
When we compare the achievable rate expressions in \eqref{eq:achievable_rate_common} and \eqref{eq:achievable_rate_unicast} with the MMSE expressions in \eqref{eq:mmse_common} and \eqref{eq:mmse_unicast}, we can obtain the following relation
\begin{align}
\mathsf{R}_{\mathsf{c},n}^\mathsf{t} &= - \log\left( \frac{\varepsilon_{\mathsf{c},n}^\mathsf{t,min}}{\mathsf{B}^\mathsf{t} }\right), \label{eq:rate_mmse_common}\\
\mathsf{R}_{\mathsf{u},n}^\mathsf{t} &= -\log \left( \frac {\varepsilon_{\mathsf{u},n}^\mathsf{t,min}}{\mathsf{C}^\mathsf{t}}\right). \label{eq:rate_mmse_unicast}
\end{align}
Considering the relations given by \eqref{eq:rate_mmse_common} and \eqref{eq:rate_mmse_unicast}, we define the WMSE expressions for the common and unicast messages, respectively, as follows \cite{Christensen2008_WeightedSumRate}
\begin{align}
\xi_{\mathsf{c},n}^\mathsf{t} &= w_n^\mathsf{t} \, \varepsilon_{\mathsf{c},n}^\mathsf{t} \,{-}\, \log \! \left( w_n^\mathsf{t} \, \mathsf{B}^\mathsf{t} \right) ,\label{eq:wmse_common}  \\
\xi_{\mathsf{u},n}^\mathsf{t} &= v_n^\mathsf{t} \, \varepsilon_{\mathsf{u},n}^\mathsf{t} \,{-}\, \log \! \left( v_n^\mathsf{t} \, \mathsf{C}^\mathsf{t} \right) ,\label{eq:wmse_unicast}
\end{align} 
where $w_n^\mathsf{t}$ and $v_n^\mathsf{t}$ are the nonzero weight coefficients of the WMMSE receiver for the common and unicast messages, respectively, of the $n$-th user. Assuming that $\xi_{\mathsf{c},n}^\mathsf{t,min}$ and $\xi_{\mathsf{u},n}^\mathsf{t,min}$ are the minimum of \eqref{eq:wmse_common} and \eqref{eq:wmse_unicast}, respectively, over all possible the WMMSE receivers and weights, we describe the respective optimization problem as follows
\begin{IEEEeqnarray}{rl}
\min_{\textbf{F}^\mathsf{t}, \textbf{G}^\mathsf{t}}
&\qquad \max_{k \,{\in}\, \mathcal{K}} \; \max_{n \,{\in}\, \mathcal{G}_k}   \; \xi_{\mathsf{u},n}^\mathsf{t,min} \label{eq:optimization_MMFWMMSE1} \\
\text{s.t.}
&\qquad \xi_{\mathsf{c},n}^\mathsf{t,min} \leq \xi_\mathsf{c}^\mathsf{th}, \, \forall n \,{\in}\, \mathcal{N} , \IEEEyessubnumber \label{eq:optimization_WMMSE_common_rate}\\
&\qquad \eqref{pow_const_tx},  \eqref{pow_const_rel},
\IEEEyessubnumber \label{eq:optimization_MMFWMMSE_power}
\end{IEEEeqnarray}
which is counterpart of the rate MMF rate optimization in \eqref{eq:optimization_MMFSR1} with $\xi_\mathsf{c}^\mathsf{th}$ being the threshold MSE for the common message. Following the strategy of \eqref{eq:optimization_MMFSR2}, we reformulate \eqref{eq:optimization_MMFWMMSE1} by employing the auxiliary variable $\eta$ to obtain a smooth constrained optimization problem given as follows
\begin{IEEEeqnarray}{rl}
\min_{\textbf{F}^\mathsf{t}, \textbf{G}^\mathsf{t},\eta}
&\qquad \eta \label{eq:optimization_MMFWMMSE2} \\
\text{s.t.} & \qquad \xi_{\mathsf{u},n}^\mathsf{t,min} \leq \eta, \, \forall n \,{\in}\, \mathcal{N}, \IEEEyessubnumber \label{eq:optimization_MMFWMMSE2_multicast}\\
& \qquad \xi_{\mathsf{c},n}^\mathsf{t,min} \leq \xi_\mathsf{c}^\mathsf{th}, \, \forall n \,{\in}\, \mathcal{N} , \IEEEyessubnumber\\
&\qquad \eqref{pow_const_tx},  \eqref{pow_const_rel}.
\IEEEyessubnumber \label{eq:optimization_MMFMSE2_power}
\end{IEEEeqnarray}

\section{Iterative Precoder Design} \label{sec:precoder_design}

In this section, we first prove that the precoders maximizing MMF rate in \eqref{eq:optimization_MMFSR2} and minimizing MMF WMMSE in \eqref{eq:optimization_MMFWMMSE2} are equivalent at the optimal solution, and then propose an alternating-optimization algorithm to compute the transmit and relay precoders based on the equivalent MMF WMMSE problem.

\begin{theorem}\label{theorem:equivalence}
The optimization problem in \eqref{eq:optimization_MMFSR2}, which aims at maximizing MMF rate, is equivalent to the optimization problem in \eqref{eq:optimization_MMFWMMSE2}, which considers to minimize MMF WMMSE, if and only if the WMMSE weight coefficients for the common and the multicast messages are chosen, respectively, as follows  
\begin{align}
w_n^\mathsf{t,opt} &= \frac{1}{ \varepsilon_{\mathsf{c},n}^\mathsf{t,min}}, \label{eq:optimal_weight_common}\\
v_n^\mathsf{t,opt} &= \frac{1}{\varepsilon_{\mathsf{u},n}^\mathsf{t,min}}, \label{eq:optimal_weight_unicast}
\end{align}
where $\varepsilon_{\mathsf{u},n}^\mathsf{t,min}$ and $\varepsilon_{\mathsf{c},n}^\mathsf{t,min}$ are the MMSE values associated with the decoding of the common and unicast messages, respectively, as given by \eqref{eq:mmse_common} and \eqref{eq:mmse_unicast}, respectively. 
\end{theorem}

\color{black}
\begin{IEEEproof}
See Appendix~\ref{appx:theorem:equivalence}.
\end{IEEEproof}
\color{black}

\begin{algorithm}[!t]
    \caption{WMMSE-Based Alternating-Optimization Precoder Design}
    \label{algorithm:without_ratesplitting}
    \begin{algorithmic}[1]
        \State \textbf{Initialize:} $\alpha$, $\epsilon$, $\mathsf{P}_\mathsf{tx}$, $\mathsf{P}_\mathsf{re}$,  
        $\textbf{F}^\mathsf{t}$, $\textbf{G}^\mathsf{t}$,
        $\mathsf{R}_\mathsf{c}^\mathsf{th}$, $\xi_\mathsf{c}^\mathsf{th} \gets 1 \,{-}\, \mathsf{R}_\mathsf{c}^\mathsf{th}$,
        $l \gets 1$, $\xi_\mathsf{t}^{(0)} \gets \infty$, $\xi_\mathsf{t}^{({-}1)} \gets \infty$
        \While{$\big| \eta^{(l{-}1)} - \eta^{(l{-}2)}
        \big| > \epsilon$} 
            \State Compute $W_n^\mathsf{t}$ and $V_n^\mathsf{t}$ by \eqref{eq:mmse_receiver_common} and \eqref{eq:mmse_receiver_unicast} for given $\textbf{F}^\mathsf{t}$ and $\textbf{G}^\mathsf{t}$
            \State Compute $\varepsilon_{\mathsf{c},n}^\mathsf{t}$ and $\varepsilon_{\mathsf{u},n}^\mathsf{t}$ by \eqref{eq:mse_common} and \eqref{eq:mse_unicast} for given $\textbf{F}^\mathsf{t}$ and $\textbf{G}^\mathsf{t}$
            \State Compute $w_n^\mathsf{t,opt}$ and $v_n^\mathsf{t,opt}$ by \eqref{eq:optimal_weight_common} and \eqref{eq:optimal_weight_unicast}
            \State Update $\textbf{G}^\mathsf{t}$ by solving \eqref{eq:optimization_general} without \eqref{pow_const_tx} for given $\textbf{F}^\mathsf{t}$, $W_n^\mathsf{t}$ and $V_n^\mathsf{t}$ \label{algline:update_G}
            \State Update $\textbf{F}^\mathsf{t}$ by solving \eqref{eq:optimization_general} for given $\textbf{G}^\mathsf{t}$, $W_n^\mathsf{t}$, and $V_n^\mathsf{t}$
            \State $\eta^{(l)} \gets $ output of the optimization (\ref{eq:optimization_general})
            \State $l \gets l+1$
        \EndWhile 
    \end{algorithmic}
\end{algorithm}

We would like to note that the problem of MMF rate optimization in \eqref{eq:optimization_MMFSR2} is non-convex, and therefore requires sophisticated approaches to solve. In this work, we resort to optimizing the MMF WMMSE problem in \eqref{eq:optimization_MMFWMMSE2} to obtain a solution for \eqref{eq:optimization_MMFSR2} using the equivalency of these two problems as proved by Theorem~\ref{theorem:equivalence}. To this end, we first reformulate the optimization problem in \eqref{eq:optimization_MMFWMMSE2} by considering the WMSE expressions $\varepsilon_{\mathsf{c},n}^\mathsf{t} $ and $\varepsilon_{\mathsf{u},n}^\mathsf{t} $ (instead of the WMMSE expressions $\varepsilon_{\mathsf{c},n}^\mathsf{t,min}$ and $\varepsilon_{\mathsf{u},n}^\mathsf{t,min}$) as the objective function. By this way, we generalize the respective optimization problem for general receivers (i.e., instead of assuming WMMSE receivers), which is given by
\begin{IEEEeqnarray}{rl}
\min_{\substack{\textbf{F}^\mathsf{t}, \textbf{G}^\mathsf{t},\,\eta \\ W_n^\mathsf{t},V_n^\mathsf{t},w_n^\mathsf{t},v_n^\mathsf{t}}}
&\qquad \eta \label{eq:optimization_general} \\
\text{s.t.} & \qquad \xi_{\mathsf{u},n}^\mathsf{t} \leq \eta, \, \forall n \,{\in}\, \mathcal{N}, \IEEEyessubnumber\\
& \qquad \xi_{\mathsf{c},n}^\mathsf{t} \leq \xi_\mathsf{c}^\mathsf{th}, \, \forall n \,{\in}\, \mathcal{N} , \IEEEyessubnumber\\
&\qquad \eqref{pow_const_tx},  \eqref{pow_const_rel}.
\IEEEyessubnumber \label{eq:optimization_MMFMSE3_power}
\end{IEEEeqnarray}

The optimization problem in \eqref{eq:optimization_general} is still non-convex, and obtaining the globally optimal solution is therefore intractable. Thanks to the WMMSE formulation, we can however split this optimization problem into three convex subproblems as follows

\begin{itemize}
    \item[1--\!\!] Computing the optimal receivers $W_n^\mathsf{t}$ and $V_n^\mathsf{t}$ for given $\textbf{F}^\mathsf{t}$ and $\textbf{G}^\mathsf{t}$,
    \item[2--\!\!] Finding the optimal relay precoder $\textbf{G}^\mathsf{t}$ for given $\textbf{F}^\mathsf{t}$, $W_n^\mathsf{t}$, and $V_n^\mathsf{t}$,
    \item[3--\!\!] Calculating the optimal transmit precoder $\textbf{F}^\mathsf{t}$ for given $\textbf{G}^\mathsf{t}$, $W_n^\mathsf{t}$, $V_n^\mathsf{t}$.
\end{itemize}

As a result, we decompose the optimization problem of \eqref{eq:optimization_general} into three parts consisting of the computation of optimal receivers, optimal relay precoder, and optimal transmit precoder, and obtain the ultimate solution through alternating-optimization approach given in Algorithm~\ref{algorithm:without_ratesplitting}. Note that we do not need to consider the power constraint at the BS while updating the relay precoder (i.e., line \eqref{algline:update_G}) since the transmit precoder is treated to be given.

\begin{remark}
\label{remark:convergence}
\color{black}
We would like to note that the value of the objective function in \eqref{eq:optimization_general}, which is based on WMSE, decreases monotonically at each alternating-optimization iteration of Algorithm~\ref{algorithm:without_ratesplitting} \cite{Christensen2008_WeightedSumRate}, thanks to the corresponding MMF rates increasing monotonically, as well. Since the WMSE is lower-bounded (by the WMMSE solution), the algorithm accordingly converges \cite{Christensen2008_WeightedSumRate, Luo2011WMMSE, Utschick2006AltOpt}. We would also like to point out the discussion in \cite{Luo2011WMMSE,Utschick2006AltOpt} on the convergence of the WMMSE algorithm (based on alternating-optimization iterations) to the global optimum for a broad class of optimization problems. We, however, do not provide such a guarantee for our particular scenario.
\color{black}
% The value of the objective function increases at each iteration of the alternating-optimization procedure in Algorithm~\ref{algorithm:without_ratesplitting}, thanks to the power constraint. The proposed algorithm therefore converges to a limit value, which can be shown following the approach of \cite[Section IV-A]{Christensen2008_WeightedSumRate} and  \cite{Kaleva2016_DecentralizedSumRateMaximization}. This limit value is, however, not guaranteed to be the global optimum, which is due to the non-convexity of the problem. With that said, since the overall algorithm employs the precoders and the MMSE receivers of Theorem~\ref{theorem:equivalence}, all of which satisfy the KKT conditions of the underlying WMMSE problem, the respective solution is at least a locally optimum with satisfactory performance, as is depicted in Section~\ref{sec:results}.
\end{remark}

\section{Precoder Design for Rate Splitting} 
\label{sec:rate_splitting}

In this section, we consider rate splitting (RS) strategy to transmit the desired common and multicast message streams to the intended users. To this end, we first consider the modified transmission model and respective achievable rates, and then present the equivalent WMMSE problem and corresponding precoder design.  

\subsection{Transmission Model and Achievable Rates for RS}
\label{subsec:rate_splitting_model}

In the proposed RS strategy, we assume that the multicast message stream of the $k$-th group is split into \textit{common} and \textit{private} parts. The common parts of all the users' multicast message stream are then encoded into a super-common message $s_\mathsf{sc}$ together with the system-wide common message $s_\mathsf{c}$. In addition, the private part of the $k$-th group's multicast message stream is independently encoded into the message $s_{\mathsf{up},k}$ for $k \,{\in}\, \mathcal{K}$. We note that the super-common message of the RS approach includes the common part of each user's multicast message stream, which therefore differs from the previous signal model of Section~\ref{sec:system_model} (i.e., $s_\mathsf{sc}$ is equal to $s_\mathsf{c}$ in the previous case since multicast message is not involved in the super-common message). %As before, the super-common stream $s_\mathsf{sc}$ is supposed to be decoded by each user.

The encoded message stream $\{s_\mathsf{sc},s_{\mathsf{up},1},\dots,s_{\mathsf{up},K}\}$ of the RS approach can be transmitted by using either of the SC or CC schemes, as discussed in Section~\ref{sec:system_model}. In the SC scheme, we have the overall message vector $\textbf{s}^\mathsf{RS-SC} \,{=}\, \textbf{s}^\mathsf{RS-SC}_\mathsf{c} \,{+}\, \textbf{s}^\mathsf{RS-SC}_\mathsf{u}$ where $\textbf{s}^\mathsf{RS-SC}_\mathsf{c} \,{=}\, \left[s_\mathsf{sc} \dots s_\mathsf{sc}\right]^{\rm T} {\in}\, \mathbb{C}^{K{\times}1}$ and $\textbf{s}^\mathsf{RS-SC}_\mathsf{u} \,{=}\, \left[ s_{\mathsf{up},1} \dots s_{\mathsf{up},K}\right]^{\rm T} {\in}\, \mathbb{C}^{K{\times}1}$. On the other hand, the CC scheme produces the message vector $\textbf{s}^\mathsf{RS-CC} \,{=}\, \left[s _\mathsf{sc} \, s_{\mathsf{up},1} \dots s_{\mathsf{up},K} \right]^{\rm T} {\in}\, \mathbb{C}^{(K{+}1){\times}1}$. Denoting the RS message vector in either transmission scheme as $\textbf{s}^\mathsf{t}$ such that $\mathsf{t} \,{\in}\, \{\mathsf{RS{-}SC},\mathsf{RS{-}CC}\}$, we keep rest of the transmit and receive signal models of Section~\ref{sec:system_model} the same.    

In particular, the BS transmits the following message vector in the first time slot
\begin{align}
\textbf{x}_\mathsf{T}^\mathsf{t} &= \textbf{f}_\mathsf{c}^\mathsf{\,t} s_\mathsf{sc} + \sum_{k \, {\in} \, \mathcal{K}} \textbf{f}_k^\mathsf{\,t} s_{\mathsf{up},k},
\end{align}
with the same average power constraint of \eqref{pow_const_tx}. In the second time slot, the relay forwards the received signal of the first time slot after processing it with the precoder $\textbf{G}^\mathsf{t}$ according to AF strategy, which also obeys to the average power constraint of \eqref{pow_const_rel}. The received signal at the $n$-th user is then given as
\begin{align}
 y_{\mathsf{U},n}^\mathsf{t} &= \textbf{h}_n \textbf{G}^\mathsf{t} \textbf{H}_\mathsf{SR} \textbf{f}_\mathsf{c}^\mathsf{\,t} s_\mathsf{sc} + \textbf{h}_n \textbf{G}^\mathsf{t} \textbf{H}_\mathsf{SR} \sum_{k \, {\in} \, \mathcal{K}} \textbf{f}_k^\mathsf{\,t}  s_{\mathsf{up},k} + \textbf{h}_n \textbf{G}^\mathsf{t} \textbf{n}_\mathsf{R} + n_{\mathsf{U},n}.  \label{rec_signal_2_RS}
\end{align}

As before, each user is supposed to decode the super-common message first, cancel it from its received signal, and then decode the private part of its multicast message. The achievable rates for the $n$-th user's super-common {\color{black}message} and {\color{black}private part of the} multicast messages are denoted by $\mathsf{R}_{\mathsf{sc},n}^\mathsf{t}$ and $\mathsf{R}_{\mathsf{up},n}^\mathsf{t}$, respectively, and given by the same expressions of \eqref{eq:achievable_rate_common} and \eqref{eq:achievable_rate_unicast}, respectively. Since the super-common message should be decoded by each user, the respective achievable rate is given as
\begin{align}
\mathsf{R}_{\mathsf{SC}}^\mathsf{t} &= \min_{n \in \mathcal{N}} \mathsf{R}_{\mathsf{sc},n}^\mathsf{t} = \mathsf{R}_\mathsf{C}^\mathsf{t} + \sum_{k \, {\in} \, \mathcal{K}} \mathsf{R}_{\mathsf{UC},k}^\mathsf{t} , \label{eqn:rate_supercommon_2}
\end{align}
where $\mathsf{R}_\mathsf{C}^\mathsf{t}$ and $\mathsf{R}_{\mathsf{UC},k}^\mathsf{t}$ are the constituent rates associated with the transmission of the system-wide common message $s_\mathsf{c}$ and the common part of the $k$-th group's multicast message $s_{\mathsf{uc},k}$, respectively. The optimal precoders maximizing the MMF rate is accordingly given as follows
\begin{IEEEeqnarray}{rl}
\max_{\textbf{F}^\mathsf{t}, \textbf{G}^\mathsf{t}, \theta_{\mathsf{c}}, \theta_{\mathsf{uc},k}}
&\qquad \min_{k \,{\in}\, \mathcal{K}} \left( \theta_{\mathsf{uc},k}+ \min_{n \,{\in}\, \mathcal{G}_k} \mathsf{R}_{\mathsf{up},n}^\mathsf{t} \right) \label{eq:optimization_RS1} \\
\text{s.t.} &\qquad \theta_{\mathsf{c}} + \sum_{k \, {\in} \, \mathcal{K}} \theta_{\mathsf{uc},k} \leq \mathsf{R}_{\mathsf{sc},n}^\mathsf{t}, \, \forall n \,{\in}\, \mathcal{N} , \IEEEyessubnumber\\
&\qquad  0 \leq \theta_{\mathsf{uc},k}, \, \forall k \,{\in}\, \mathcal{K} , \IEEEyessubnumber\\
&\qquad \mathsf{R}_\mathsf{c}^\mathsf{th} \leq \theta_\mathsf{c} , \IEEEyessubnumber\\
&\qquad \eqref{pow_const_tx},  \eqref{pow_const_rel},
\IEEEyessubnumber \label{eq:optimization_RS1_power}
\end{IEEEeqnarray}
where \eqref{eq:optimization_RS1_power} represents the power constraints at the transmitter and relay ends. The optimization problem in \eqref{eq:optimization_RS1} is non-convex, as before, and is converted to a smooth constrained optimization problem by defining a new auxiliary $\Theta_k$ and $\theta_\mathsf{g}$ as follows
\begin{IEEEeqnarray}{rl}
\max_{\substack{\textbf{F}^\mathsf{t}, \textbf{G}^\mathsf{t}, \theta_{\mathsf{uc},k},\\ \Theta_k, \theta_\mathsf{g}, \theta_{\mathsf{c}}}}
&\qquad \theta_\mathsf{g} \label{eq:optimization_RS2} \\
\text{s.t.} &\qquad  \theta_\mathsf{g} \leq  \theta_{\mathsf{uc},k}+ \Theta_k, \, \forall k \, {\in} \, \mathcal{K} , \IEEEyessubnumber\\
& \qquad \Theta_k \leq \mathsf{R}_{\mathsf{up},n}^\mathsf{t}, \, \forall n \in \mathcal{G}_k, \, \forall k \, {\in} \, \mathcal{K} , \IEEEyessubnumber\\
&\qquad \theta_{\mathsf{c}} + \sum_{k \, {\in} \, \mathcal{K}} \theta_{\mathsf{uc},k} \leq \mathsf{R}_{\mathsf{sc},n}^\mathsf{t}, \, \forall n \,{\in}\, \mathcal{N} , \IEEEyessubnumber\\
&\qquad  0 \leq \theta_{\mathsf{uc},k}, \, \forall k \,{\in}\, \mathcal{K} , \IEEEyessubnumber\\
&\qquad \mathsf{R}_\mathsf{c}^\mathsf{th} \leq \theta_\mathsf{c}, \IEEEyessubnumber\\
&\qquad \eqref{pow_const_tx},  \eqref{pow_const_rel}.
\IEEEyessubnumber \label{eq:optimization_RS2_power}
\end{IEEEeqnarray}

\subsection{WMMSE-Based Precoder Design for RS}

Similar to the strategy of Section~\ref{sec:wmmse}, we start constructing the equivalent WMMSE problem by considering the receivers $W_n^\mathsf{t}$ and $V_n^\mathsf{t}$, which correspond to the super-common message and private part of the unicast message for the $n$-th user. Assuming perfect SIC, these receivers yield the desired estimates $\hat{s}_{\mathsf{sc},n}^\mathsf{t} \,{=}\, W_n^\mathsf{t} y_{\mathsf{U},n}^\mathsf{t}$ and $\hat{s}_{\mathsf{up},n}^\mathsf{t} \,{=}\, V_n^\mathsf{t} \left( y_{\mathsf{U},n}^\mathsf{t} - \textbf{h}_n \textbf{G}^\mathsf{t} \textbf{H}_\mathsf{SR} \textbf{f}_\mathsf{c}^\mathsf{\,t} s_\mathsf{sc} \right)$. The corresponding MSE expressions $\varepsilon_{\mathsf{sc},n}^\mathsf{t}$ and $\varepsilon_{\mathsf{up},n}^\mathsf{t}$ are given by \eqref{eq:mse_common} and \eqref{eq:mse_unicast}, respectively. The optimal values of these receivers $W_n^\mathsf{t,opt}$ and $V_n^\mathsf{t,opt}$ can be found by \eqref{eq:mmse_receiver_common} and \eqref{eq:mmse_receiver_unicast}, respectively, and the corresponding MMSE expressions $\varepsilon_{\mathsf{sc},n}^\mathsf{t,min}$ and $\varepsilon_{\mathsf{up},n}^\mathsf{t,min}$ are given by \eqref{eq:mmse_common} and \eqref{eq:mmse_unicast}, respectively. 

The corresponding WMSE expressions can be set up similar to \eqref{eq:wmse_common}-\eqref{eq:wmse_unicast} as follows 
\begin{align}
\xi_{\mathsf{sc},n}^\mathsf{t} &= w_n^\mathsf{t} \varepsilon_{\mathsf{sc},n}^\mathsf{t} - \log (\mathsf{B}^\mathsf{t} w_n^\mathsf{t}) ,\label{eq:wmse_common_RS}  \\
\xi_{\mathsf{up},n}^\mathsf{t} &= v_n^\mathsf{t} \varepsilon_{\mathsf{up},n}^\mathsf{t} - \log (\mathsf{C}^\mathsf{t} v_n^\mathsf{t}) ,\label{eq:wmse_unicast_RS}
\end{align} 
for which $W_n^\mathsf{t,opt}$ and $V_n^\mathsf{t,opt}$ in \eqref{eq:mmse_receiver_common} and \eqref{eq:mmse_receiver_unicast} still provide the optimal solutions. Note that the optimal precoders $\textbf{F}^\mathsf{t \, \star}$ and $\textbf{G}^\mathsf{t \, \star}$ minimizes the MMF WMSEs of \eqref{eq:wmse_common_RS} and \eqref{eq:wmse_unicast_RS} over all the users and groups. In addition, following the strategy of Theorem~\ref{theorem:equivalence}, we can show that these optimal precoders also maximize the MMF rate in \eqref{eq:optimization_RS2} as long as the weight coefficients satisfy the following conditions as
\begin{align}
w_n^\mathsf{t,opt} &= \frac{1}{ \varepsilon_{\mathsf{sc},n}^\mathsf{t,min}}, \label{eq:optimal_weight_common_RS}
\end{align}
\begin{align}
v_n^\mathsf{t,opt} &= \frac{1}{\varepsilon_{\mathsf{up},n}^\mathsf{t,min}}. \label{eq:optimal_weight_unicast_RS}
\end{align}At this equality condition, the respective WMMSE expressions are obtained as  $\xi_{\mathsf{sc},n}^\mathsf{t,min} \,{=}\, 1 \,{-}\, \mathsf{R}_{\mathsf{sc},n}^\mathsf{t}$ and $\xi_{\mathsf{up},n}^\mathsf{t,min} \,{=}\, 1 \,{-}\, \mathsf{R}_{\mathsf{up},n}^\mathsf{t}$, and we have $\xi_\mathsf{c}^\mathsf{th} = 1- \mathsf{R}_\mathsf{c}^\mathsf{th}$ as before. The optimization problem in \eqref{eq:optimization_RS2} can be expressed as follows
\begin{IEEEeqnarray}{rl}
\min_{\substack{\textbf{F}^\mathsf{t}, \textbf{G}^\mathsf{t}, \eta_{\mathsf{uc},k},\\ \Gamma_k, \eta_\mathsf{g}, \eta_{\mathsf{c}}}}
&\qquad \eta_\mathsf{g} \label{eq:optimization_WSR_RS_equivalent} \\
\text{s.t.} &\qquad  \eta_{\mathsf{uc},k}+ \Gamma_k \leq \eta_\mathsf{g}, \, \forall k \, {\in} \, \mathcal{K} , \IEEEyessubnumber\\
& \qquad  \xi_{\mathsf{up},n}^\mathsf{t,min} \leq \Gamma_k, \, \forall n \in \mathcal{G}_k, \, \forall k \, {\in} \, \mathcal{K} , \IEEEyessubnumber\\
&\qquad  \xi_{\mathsf{sc},n}^\mathsf{t,min} \leq \eta_{\mathsf{c}} + \sum_{k \, {\in} \, \mathcal{K}} \eta_{\mathsf{uc},k}, \, \forall n \,{\in}\, \mathcal{N} , \IEEEyessubnumber\\
&\qquad \eta_{\mathsf{uc},k} \leq 1, \, \forall k \,{\in}\, \mathcal{K} , \IEEEyessubnumber\\
&\qquad \eta_\mathsf{c} \leq \xi_\mathsf{c}^\mathsf{th} , \IEEEyessubnumber\\
&\qquad \eqref{pow_const_tx},  \eqref{pow_const_rel}.
\IEEEyessubnumber \label{eq:optimization_RS3_power}
\end{IEEEeqnarray}

\begin{algorithm}[!t]
    \caption{Alternating-Optimization Precoder Design for Rate Splitting}
    \label{algorithm:with_ratesplitting}
    \begin{algorithmic}[1]
        \State \textbf{Initialize:} $\alpha$, $\epsilon$, $\mathsf{P}_\mathsf{tx}$, $\mathsf{P}_\mathsf{re}$,  
        $\textbf{F}^\mathsf{t}$, $\textbf{G}^\mathsf{t}$,
        $\mathsf{R}_\mathsf{c}^\mathsf{th}$, $\eta_\mathsf{g}^\mathsf{th} \gets 1 \,{-}\, \mathsf{R}_\mathsf{c}^\mathsf{th}$,
        $l \gets 1$, $\xi_\mathsf{t}^{(0)} \gets \infty$, $\xi_\mathsf{t}^{({-}1)} \gets \infty$
        \While{$\big| \eta_\mathsf{g}^{(l{-}1)} - \eta_\mathsf{g}^{(l{-}2)}
        \big| > \epsilon$} 
            \State Compute $W_n^\mathsf{t}$ and $V_n^\mathsf{t}$ by \eqref{eq:mmse_receiver_common} and \eqref{eq:mmse_receiver_unicast} for given $\textbf{F}^\mathsf{t}$ and $\textbf{G}^\mathsf{t}$
            \State Compute $\varepsilon_{\mathsf{sc},n}^\mathsf{t,min}$ and $\varepsilon_{\mathsf{up},n}^\mathsf{t,min}$ by \eqref{eq:mmse_common} and \eqref{eq:mmse_unicast} for given $\textbf{F}^\mathsf{t}$ and $\textbf{G}^\mathsf{t}$
            \State Compute $w_n^\mathsf{t,opt}$ and $v_n^\mathsf{t,opt}$ by \eqref{eq:optimal_weight_common_RS} and \eqref{eq:optimal_weight_unicast_RS}
            \State Update $\textbf{G}^\mathsf{t}$ by solving \eqref{eq:optimization_WSR_RS_equivalent} without \eqref{pow_const_tx} for given $\textbf{F}^\mathsf{t}$, $W_n^\mathsf{t}$ and $V_n^\mathsf{t}$
            \State Update $\textbf{F}^\mathsf{t}$ by solving \eqref{eq:optimization_WSR_RS_equivalent} for given $\textbf{G}^\mathsf{t}$, $W_n^\mathsf{t}$ and $V_n^\mathsf{t}$
            \State $\eta_\mathsf{g}^{(l)} \gets$ output of the optimization (\ref{eq:optimization_WSR_RS_equivalent})
            \State $l \gets l+1$
        \EndWhile 
    \end{algorithmic}
\end{algorithm}

Although the optimization problem in \eqref{eq:optimization_WSR_RS_equivalent} is non-convex in the joint set of optimization variables $\left\lbrace W_n^\mathsf{t},V_n^\mathsf{t}\right\rbrace$, $\textbf{G}^\mathsf{t}$, and $\textbf{F}^\mathsf{t}$, it is convex in each of these variables while keeping the others fixed. We therefore consider to optimize each of these variables separately through alternating-optimization approach, which is detailed in Algorithm~\ref{algorithm:with_ratesplitting}.

\section{Simulation Results} \label{sec:results}

In this section, we present numerical results based on extensive Monte Carlo simulations to evaluate the performance of the proposed transmission strategies (i.e., SC, CC, RS-CC, and RS-CC). In particular, we compute the transmit and relay precoders using Algorithm~\ref{algorithm:without_ratesplitting} and \ref{algorithm:with_ratesplitting}, and resort to CVX toolbox \cite{CVX} whenever necessary while solving convex optimization problems. We use optimal $\alpha$ for SC and RS-SC schemes, and present rates in terms of bits per channel use (BPCU). {\color{black}In addition, we use all-one matrices while initializing the transmit and relay precoders unless otherwise stated.}

\begin{figure}[!t]
\centering
\includegraphics[width=0.715\textwidth]{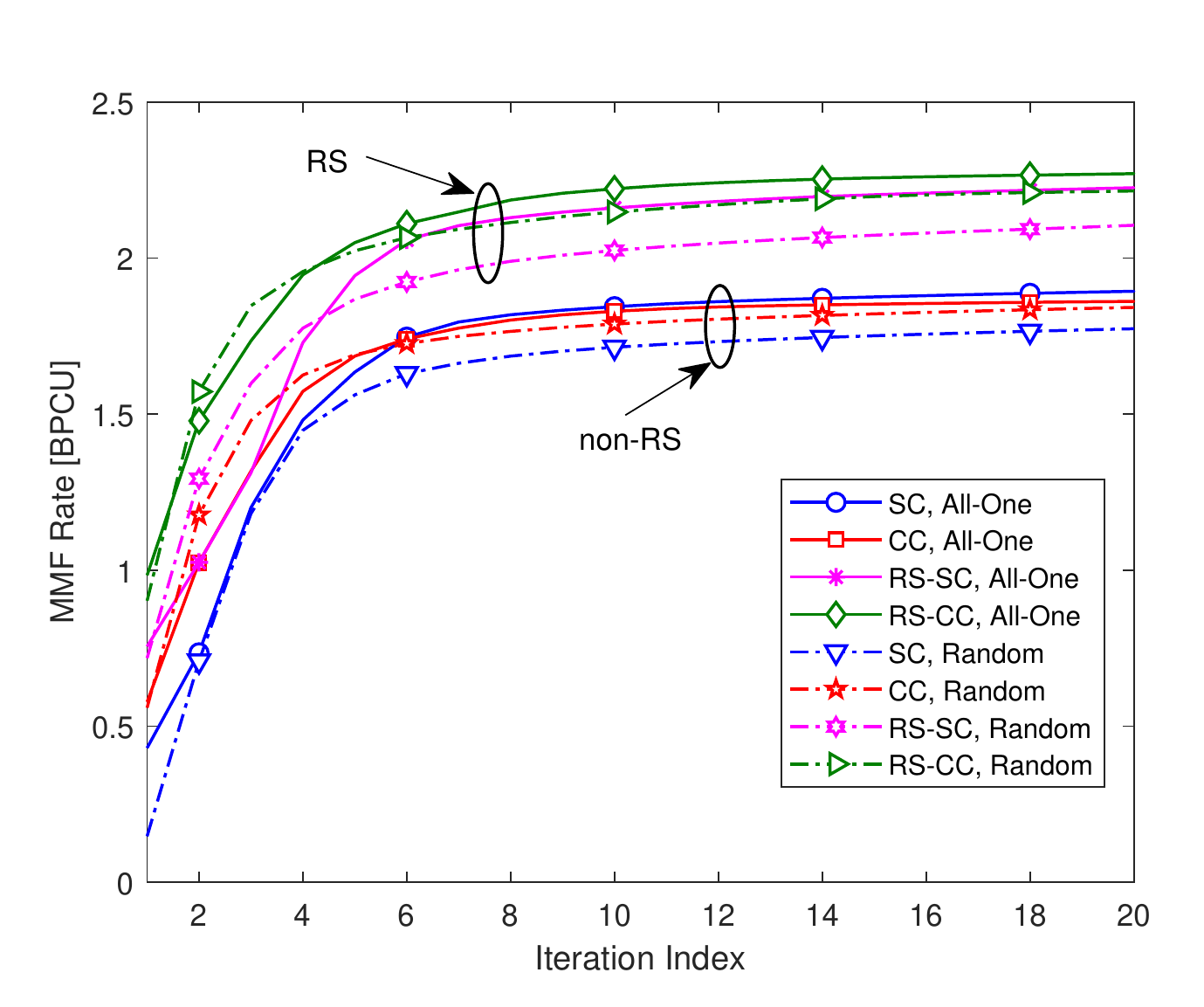}
\caption{Convergence of MMF rates {\color{black}assuming the precoder initialization with all-one and random matrices} for $M\,{=}\,3$, $N_\mathsf{R}\,{=}\,3$, $N\,{=}\,K\,{=}\,3$, $(G_1,G_2,G_3) \,{=}\,(1,1,1)$, and $\mathsf{R}_\mathsf{c}^\mathsf{th}\,{=}\,0.5$ BPCU at transmit SNR of $\gamma \,{=}\, 15\,\text{dB}$.} \label{fig:convergence}
\centering
\end{figure}

In Fig.~\ref{fig:convergence}, we depict the convergence rate of the alternating-optimization schemes given by Algorithm~\ref{algorithm:without_ratesplitting} and \ref{algorithm:with_ratesplitting} for {\color{black}the precoder initialization schemes employing all-one matrix as well as random matrix (with standard complex Gaussian entries)}. In particular, we assume $M\,{=}\,3$ and $N_\mathsf{R}\,{=}\,3$ for the number of antennas at the BS and relay, respectively, $N\,{=}\,K\,{=}\,3$ for the number of users and multicast groups (i.e., each group has a single user), common rate threshold of  $\mathsf{R}_\mathsf{c}^\mathsf{th}\,{=}\,0.5$ BPCU, and transmit SNR of $\gamma \,{=}\, 15\,\text{dB}$. We observe that all the transmission strategies converge very quickly. We also observe that {\color{black}while SC (RS-SC) and CC (RS-CC) under the precoder initialization with all-one matrix converge to very similar MMF rates, random initialization deteriorates the rate performance making CC (RS-CC) superior to SC (RS-SC). In addition, for both initialization schemes, the convergence speed of CC (RS-CC) is better than that of SC (RS-SC). Both these observations point out the superiority of the CC scheme to the SC scheme in terms of convergence rate for both non-RS and RS settings.}

\begin{figure}[!t]
\centering
\includegraphics[width=0.715\textwidth]{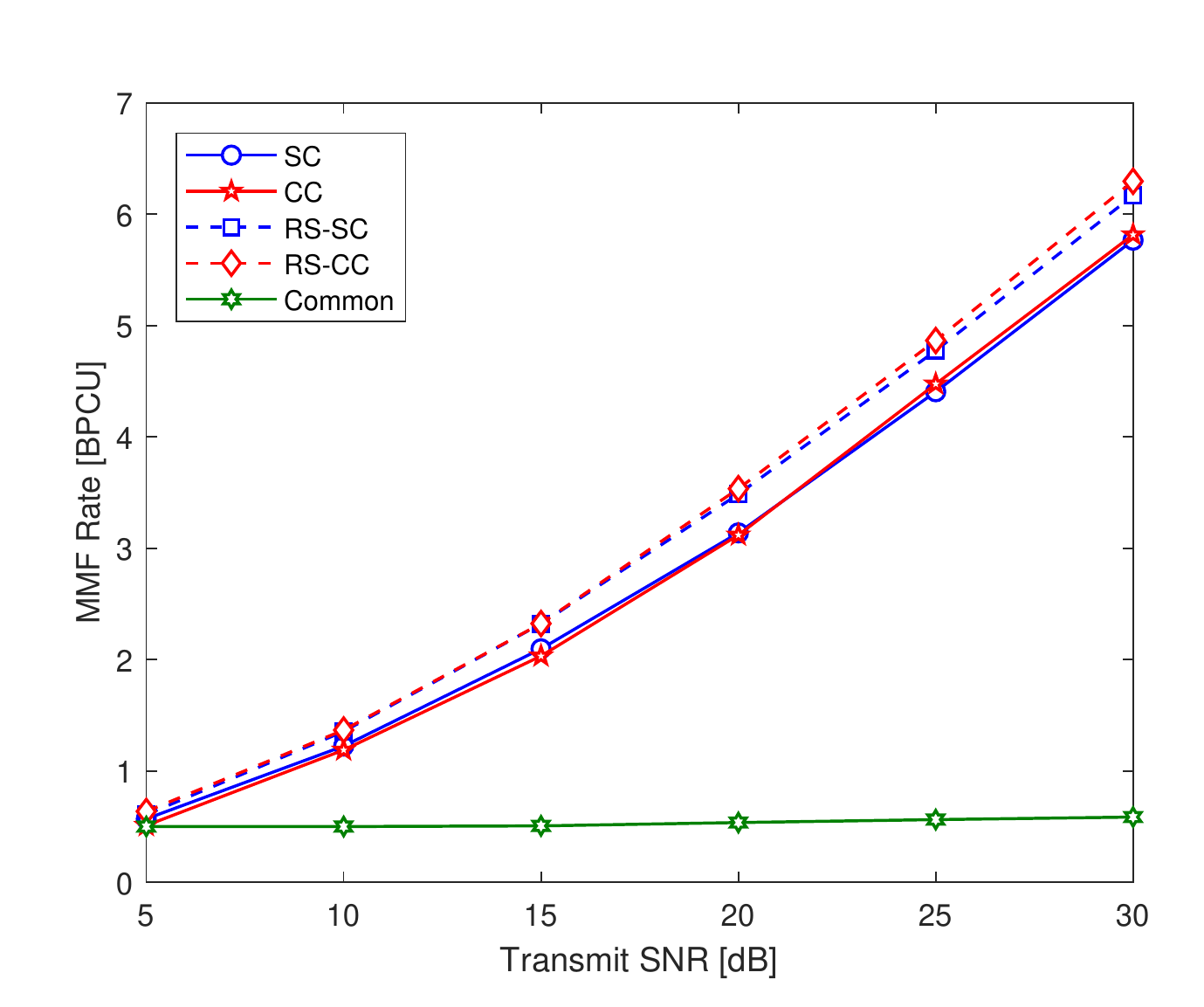}
\caption{MMF and common message rates for $M\,{=}\,3$, $N_\mathsf{R}\,{=}\,3$, $N\,{=}\,K\,{=}\,3$, $(G_1,G_2,G_3) \,{=}\,(1,1,1)$, and $\mathsf{R}_\mathsf{c}^\mathsf{th}\,{=}\,0.5$ BPCU.} \label{fig:M3N3K111}
\end{figure}

In Fig.~\ref{fig:M3N3K111}, we depict the MMF and common message rates of all the transmission strategies under consideration keeping the communications setting of Fig.~\ref{fig:convergence} the same. We observe that RS-based schemes (i.e., RS-SC and RS-CC) outperform non-RS schemes (i.e., SC and CC). In addition, there is almost no difference between the SC and CC schemes without RS while CC scheme slightly outperforms SC when RS is applied. We also observe that the common rate threshold is successfully met at each transmit SNR level. Note that multicast groups in this example are composed of a single user, and hence do not fully exploit the performance of the MMF design in general.% In the following examples, we look into the impact of multicast group size on the user rates under various settings. 

\begin{figure}[!t]
\centering
\includegraphics[width=0.715\textwidth]{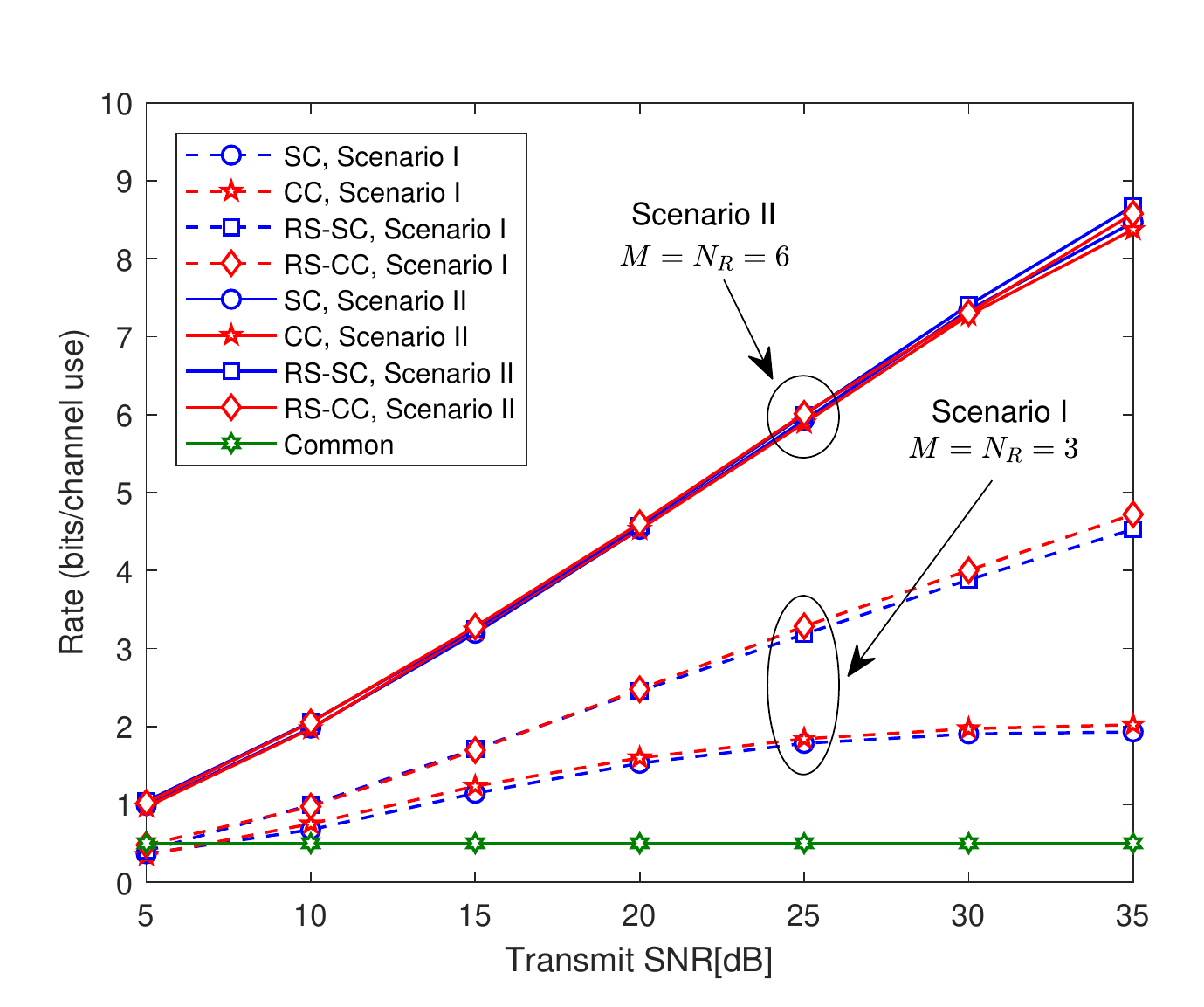}
\caption{MMF and common message rates for $(M,N_\mathsf{R})\,{=}\,\{(3,3),(6,6)\}$, $N\,{=}\,6$, $K\,{=}\,3$, $(G_1,G_2,G_3) \,{=}\,(1,2,3)$, and $\mathsf{R}_\mathsf{c}^\mathsf{th}\,{=}\,0.5$ BPCU.}\label{fig:M36N36K123}
\end{figure}

In Fig.~\ref{fig:M36N36K123}, we investigate the impact of multicast-group size along with the number of users and antennas. In particular, we assume $K\,{=}\,3$ multicast groups having $(G_1,G_2,G_3) \,{=}\,(1,2,3)$ users (i.e., $N\,{=}\,6$ users in total), the number of antenna pairs being $(M,N_\mathsf{R})\,{=}\,\{(3,3),(6,6)\}$, and common rate threshold of $\mathsf{R}_\mathsf{c}^\mathsf{th}\,{=}\,0.5$ BPCU. We observe that RS-based schemes perform much better than non-RS schemes for $(M,N_\mathsf{R})\,{=}\,(3,3)$ while applying RS does not produce any improvement on the rates for $(M,N_\mathsf{R})\,{=}\,(6,6)$. As before, there is almost no difference between the SC and CC schemes without RS, and RS-CC slightly outperforms RS-SC in general. We also note that RS-based schemes do not saturate with increasing transmit SNR while non-RS schemes saturate for $(M,N_\mathsf{R})\,{=}\,(3,3)$ having relatively higher multiuser interference.

\begin{figure}[!t]
\centering
\includegraphics[width=0.715\textwidth]{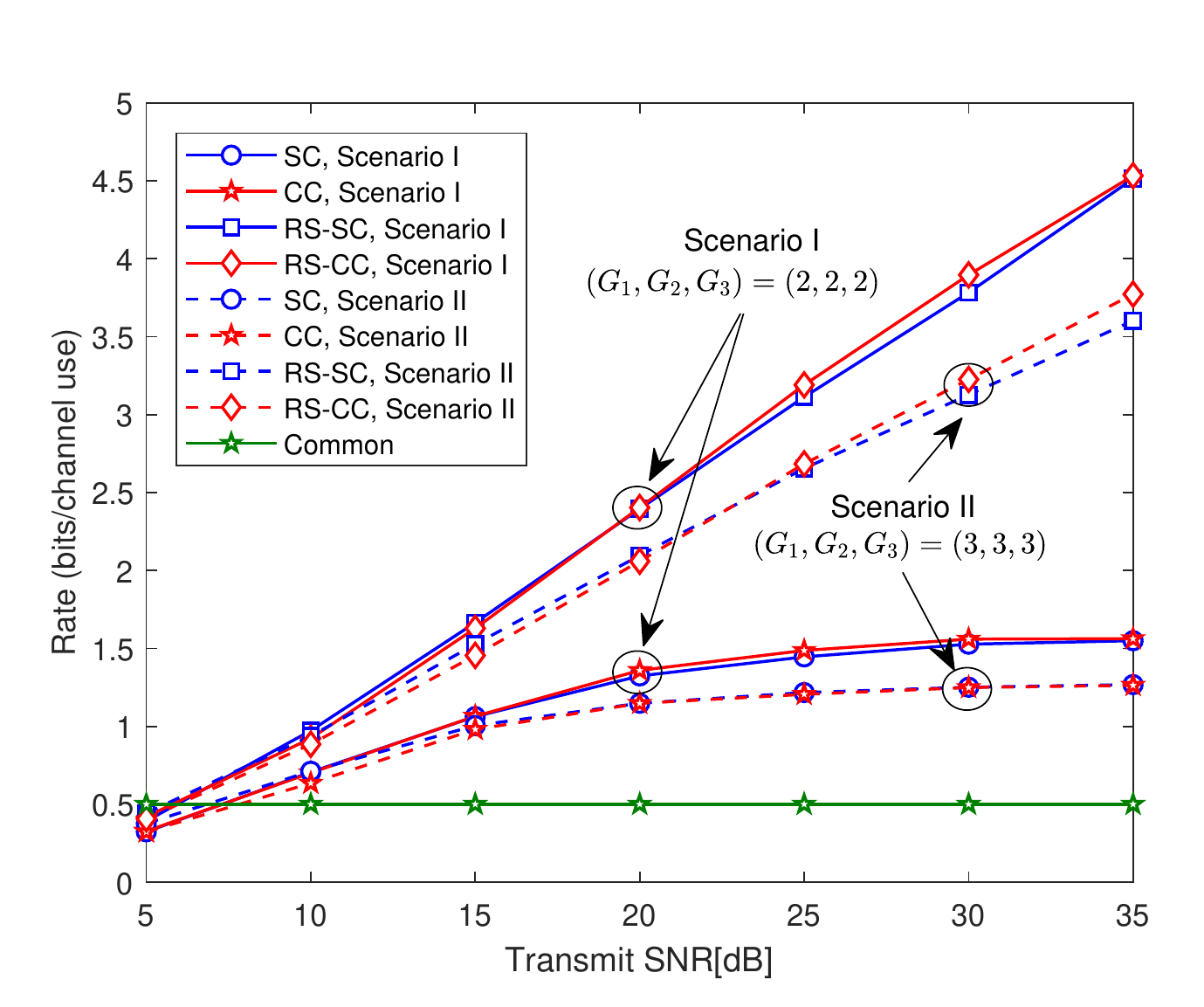}
\caption{MMF and common rates for $M\,{=}\,3$, $N_\mathsf{R}\,{=}\,3$, $N\,{=}\,6$, $K\,{=}\,3$, {\color{black} $(G_1,G_2,G_3) \,{=}\,\{(2,2,2),(3,3,3)\}$}, and $\mathsf{R}_\mathsf{c}^\mathsf{th}\,{=}\,0.5$ BPCU.} \label{fig:M3N3K222}
\end{figure}

\begin{remark} \label{remark:overloading}
We would like to recall that the RS approach offers to decode the multiuser interference \textit{partially} (as opposed to NOMA trying to decode all the interference for the strong user), and hence is expected to manifest its strength under high multiuser interference scenarios. This discussion aligns with the outcomes of Fig.~\ref{fig:M36N36K123} since $(M,N_\mathsf{R})\,{=}\,(3,3)$ setting corresponds to an \textit{overloaded} scenario with more severe multiuser interference (i.e., $3{\times}3$ antennas serving $6$ users), and RS-based schemes therefore perform much better than non-RS schemes. In contrast, the multiuser interference is not that severe for $(M,N_\mathsf{R})\,{=}\,(6,6)$ (i.e., $6{\times}6$ antennas serving $6$ users), and RS does not portray any additional improvement. To  better investigate this issue, we provide MMF rates for $(M,N_\mathsf{R})\,{=}\,(3,3)$ case in Fig.~\ref{fig:M3N3K222} assuming $K\,{=}\,3$ multicast groups with $N\,{=}\,\{6,9\}$ users which are evenly distributed over the groups, i.e., $(G_1,G_2,G_3) \,{=}\,\{(2,2,2),(3,3,3)\}$. We observe that RS-based schemes significantly outperform non-RS schemes for these highly overloaded scenarios, where non-RS schemes are observed to saturate after a modest transmit SNR {\color{black}due to the system performance becoming \textit{interference-limited}}.
\end{remark}

\begin{figure}[!t]
\centering
\includegraphics[width=0.715\textwidth]{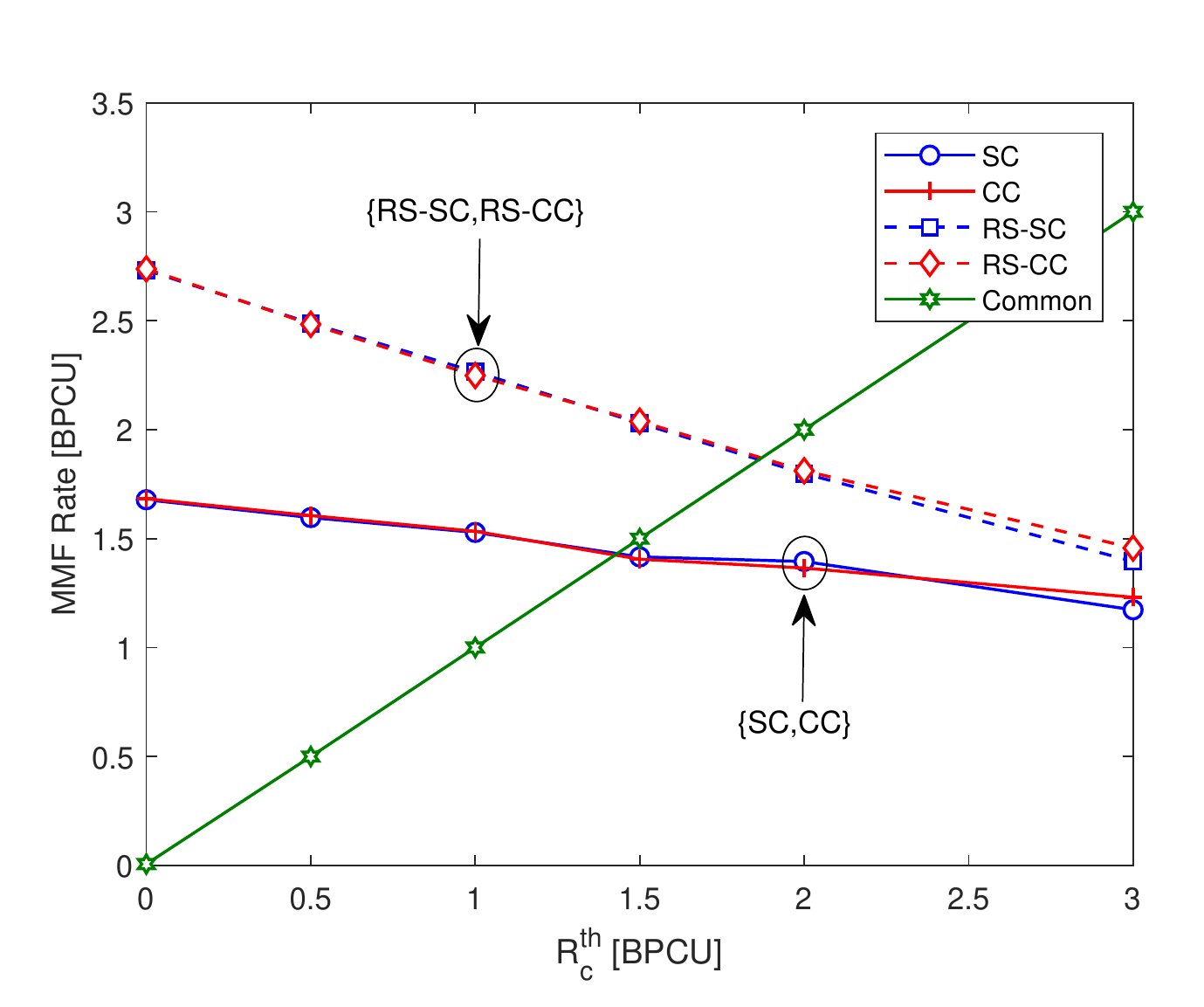}
\caption{MMF and common rates versus along with common rate threshold for $M\,{=}\,3$, $N_\mathsf{R}\,{=}\,3$, $N\,{=}\,6$, $K\,{=}\,3$, and $(G_1,G_2,G_3) \,{=}\,(1,2,3)$ at transmit SNR of $\gamma\,{=}\,20\,\text{dB}$.} \label{fig:Rcth_vs_rate}
\end{figure}
\begin{figure}[!t]
\centering
\includegraphics[width=0.715\textwidth]{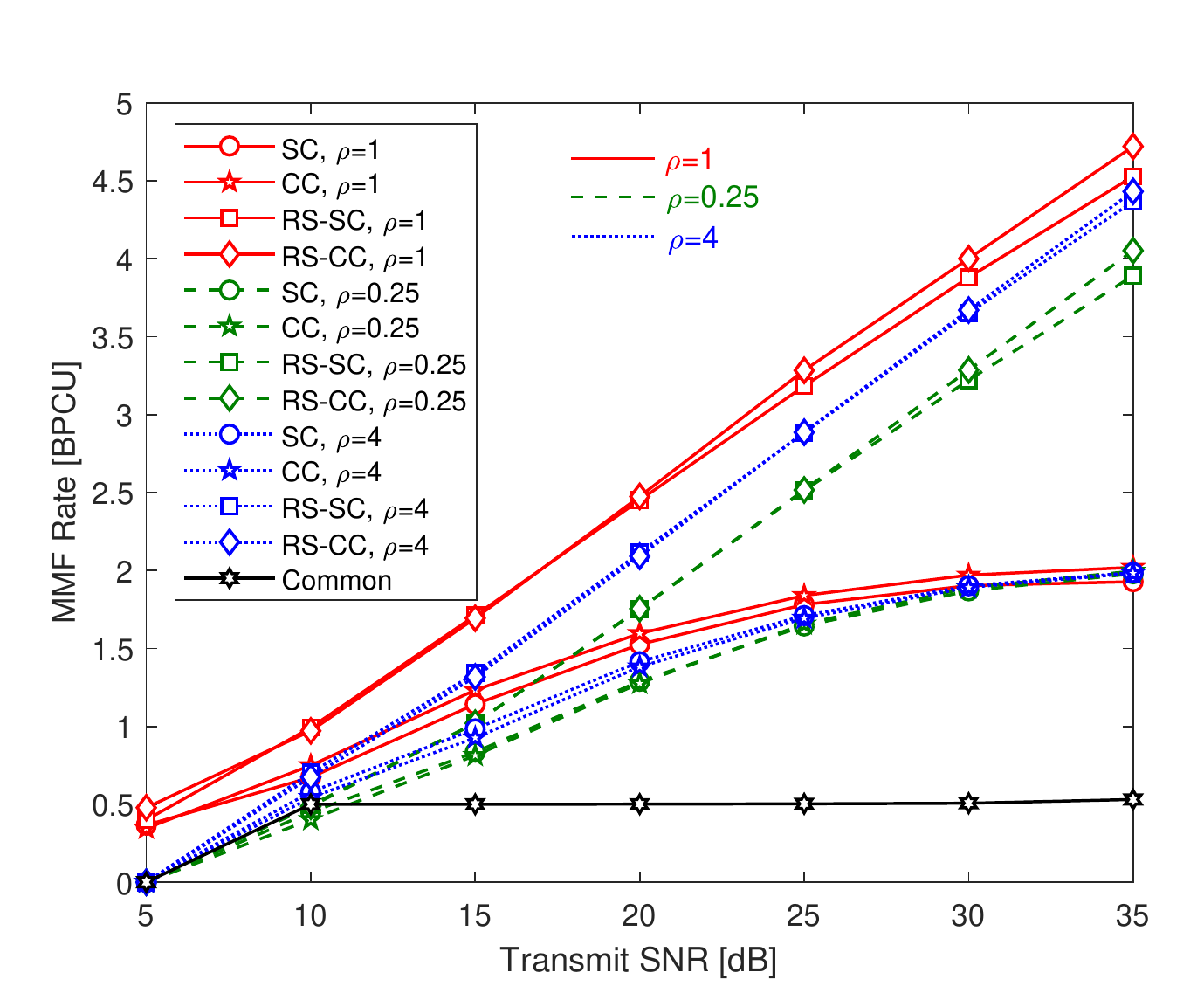}
\caption{MMF and common rates for different relay locations represented by $\rho \,{=}\, d_\mathsf{SR}/d_\mathsf{RD} \,{=}\, \{0.25,1,4\}$ for $M\,{=}\,3$, $N_\mathsf{R}\,{=}\,3$, {\color{black} $N\,{=}\,6$, $K\,{=}\,3$}, and $(G_1,G_2,G_3) \,{=}\,(1,2,3)$.} \label{fig:Pathloss_vs_rate}
\end{figure}

% \begin{figure}[!t]
% \centering
% \includegraphics[width=0.715\textwidth]{imperfectCSI.eps}
% \caption{MMF rates for different imperfect CSI parameters for $M\,{=}\,3$, $N_\mathsf{R}\,{=}\,3$, $N\,{=}\,\{3,6\}$, $K\,{=}\,3$, and $(G_1,G_2,G_3) \,{=}\,(1,1,1), (1,2,3)$.} \label{fig:Pathloss_vs_rate_imperfect}
% \end{figure}
In Fig.~\ref{fig:Rcth_vs_rate}, we look into the MMF rate performance of the transmission schemes under consideration along with varying common rate threshold $\mathsf{R}_\mathsf{c}^\mathsf{th}$. To this end, we assume a scenario with the number of antennas being $M\,{=}\,3$ and $N_\mathsf{R}\,{=}\,3$, where a total of $N\,{=}\,6$ users are distributed over $K\,{=}\,3$ multicast group unevenly, i.e., $(G_1,G_2,G_3) \,{=}\,(1,2,3)$, and transmit SNR is $\gamma\,{=}\,20\,\text{dB}$. As expected, MMF rates decrease along with increasing common rate threshold. We also observe that for any common rate threshold, RS-based transmission schemes significantly outperform non-RS schemes. Note that when there is no common message to broadcast (i.e., $\mathsf{R}_\mathsf{c}^\mathsf{th}\,{=}\,0$), RS-based schemes achieve an MMF rate that is roughly twice as large as that of non-RS schemes. As a final remark, performance gap between RS-based and non-RS schemes diminish as common rate threshold increases since---basically--degrees of freedom available for achieving MMF rates also decreases (along with increasing common rate).   

We finally investigate the impact of relay location on MMF rates. To this end, we adjust the power of the channel matrix $\textbf{H}_\mathsf{SR}$ and channel vector $\textbf{h}_k$ by $1/d_\mathsf{SR}^2$ and $1/d_\mathsf{RD}^2$, respectively, where where $d_\mathsf{SR}$ and $d_\mathsf{RD}$ are the \textit{relative} distances between the BS and relay, and relay and each user, respectively, with the ratio $\rho \,{=}\, d_\mathsf{SR}/d_\mathsf{RD}$. In particular, we adopt the setting of Fig.~\ref{fig:M36N36K123} with $M\,{=}\,3$ and $N_\mathsf{R}\,{=}\,3$, and plot MMF rates in Fig.~\ref{fig:Pathloss_vs_rate} for $\rho \,{=}\, \{0.25,1,4\}$. We observe that the mid-position for the relay (i.e., $\rho \,{=}\, 1$) achieves the best performance while the relay location being close to users (i.e., $\rho \,{=}\, 4$) is more advantageous than that being close to the BS (i.e., $\rho \,{=}\, 0.25$). We also note that the performance gap between RS-based schemes for various relay locations stay the same as transmit SNR increases while non-RS schemes saturate at the same MMF rate value SNR irrespective of the relay location along with transmit SNR getting larger.    

\section{Conclusion} \label{sec:conclusion}
We consider multigroup multicast transmission with a common message in a cooperative point-to-multipoint communications scenario. We propose superposition and concatenated coding schemes with and without RS, and derive low-complexity alternating-optimization algorithm for each transmission scheme to maximize MMF rates. We rigorously evaluate the performance of the proposed schemes, and observe that RS-based schemes are superior to non-RS ones in terms of MMF rates, especially in overloaded scenarios, without any saturation with increasing SNR.

\begin{appendices}

\color{black}
\section{} \label{appx:theorem:equivalence} 
In order to prove the equivalency of the optimization problems in \eqref{eq:optimization_MMFSR2} and \eqref{eq:optimization_MMFWMMSE2}, we consider the respective Lagrangian expressions, which is given for \eqref{eq:optimization_MMFSR2} as follows
\begin{align}
f\left(\textbf{F}^\mathsf{t},\textbf{G}^\mathsf{t},\theta \right) &= -  \theta + \sum_{n \in \mathcal{N}} \nu_n^\mathsf{t}\left(\theta - \mathsf{R}_{\mathsf{u},n}^\mathsf{t} \right) + \sum_{n \in \mathcal{N}} \kappa_n^\mathsf{t} \left(\mathsf{R}_\mathsf{c}^\mathsf{th} - \mathsf{R}_{\mathsf{c},n}^\mathsf{t}\right) + \lambda_1^\mathsf{t} J_1^\mathsf{t} + \lambda_2^\mathsf{t} J_2^\mathsf{t}, \label{eq:lagrangian_wsr}
\end{align}
where $\nu_n^\mathsf{t}$, $\kappa_n^\mathsf{t}$, $\lambda_1^\mathsf{t}$, and $\lambda_2^\mathsf{t}$ are the Lagrange multipliers, and 
\begin{align}
    J_1^\mathsf{t} &=  \mathsf{B}^\mathsf{t} \|\textbf{f}_\mathsf{c}^\mathsf{\,t}\|^2_\mathsf{F}  +  \mathsf{C}^\mathsf{t}\! \sum_{k \, {\in} \, \mathcal{K}} \! \|\textbf{f}_k^\mathsf{\,t}\|^2_\mathsf{F} - \mathsf{P}_\mathsf{tx} ,\\
    J_2^\mathsf{t} &= \mathsf{B}^\mathsf{t} \|\textbf{G}^\mathsf{t} \textbf{H}_\mathsf{SR} \textbf{f}_\mathsf{c}^\mathsf{\,t} \|^2_\mathsf{F} + \mathsf{C}^\mathsf{t}\!  \sum_{k \, {\in} \, \mathcal{K}} \!\|\textbf{G}^\mathsf{t} \textbf{H}_\mathsf{SR} \textbf{f}_k^\mathsf{\,t} \|^2_\mathsf{F} + \sigma^2 \|\textbf{G}^\mathsf{t}\|^2_\mathsf{F} - \mathsf{P}_\mathsf{re} .
\end{align}
Note that we assume the total power constraint is always satisfied with equality for this particular problem. Similarly, the Lagrangian expression for the optimization problem in \eqref{eq:optimization_MMFWMMSE2} is given as
\begin{align}
g\left(\textbf{F}^\mathsf{t},\textbf{G}^\mathsf{t},\eta \right) &= \eta + \sum_{n \in \mathcal{N}} \bar{\nu}_n^\mathsf{t} \left( \xi_{\mathsf{u},n}^\mathsf{t,min} - \eta \right) + \sum_{n \in \mathcal{N}} \bar{\kappa}_n^\mathsf{t} \left(\xi_{\mathsf{c},n}^\mathsf{t,min} - \xi_c^\mathsf{th} \right) + \bar{\lambda}_1^\mathsf{t} J_1^\mathsf{t} + \bar{\lambda}_2^\mathsf{t} J_2^\mathsf{t}, \label{eq:lagrangian_wmmse}
\end{align}
where $\bar{\nu}_n^\mathsf{t}$, $\bar{\kappa}_n^\mathsf{t}$, $\bar{\lambda}_1^\mathsf{t}$, and $\bar{\lambda}_2^\mathsf{t}$ are the Lagrange multipliers.

Since the optimization problems in \eqref{eq:optimization_MMFSR2} and \eqref{eq:optimization_MMFWMMSE2} are equivalent if the solutions to their Lagrangian expressions are the same, we calculate the gradient of \eqref{eq:lagrangian_wsr} and \eqref{eq:lagrangian_wmmse} in the following with respect to unknown precoders and auxiliary variables. Due to the space limitations, we consider only SC scheme for the derivation, and drop the superscript $\mathsf{t}$ within this proof. Defining $\Tilde{\textbf{H}}_\mathsf{SR} \,{=}\, \textbf{G} \textbf{H}_\mathsf{SR}$ and $\Tilde{\textbf{h}}_{n} \,{=}\, \textbf{h}_n \Tilde{\textbf{H}}_\mathsf{SR}$, the gradients for \eqref{eq:lagrangian_wsr} are given as follows\footnote{The gradient of a function $f(\textbf{x})$ with respect to its complex variable $\textbf{x}$ is denoted as $\nabla_{\textbf{x}} f(\textbf{x})$ with its $m$-th element being $[\nabla_{\textbf{x}}f(\textbf{x})]_{m} = \nabla_{[\textbf{x}]_{m}}f(\textbf{x}) = \frac{\partial f(\textbf{x})}{\partial[\textbf{x}^\ast]_{m}}$. For detailed derivation rules, we refer the reader to \cite{Hjorungnes2007ComVal}.}
\begin{align}
\nabla_{\textbf{f}_k} f(\textbf{F},\textbf{G},\theta) &= \sum_{i \in \mathcal{N}\setminus \mathcal{G}_k} \nu_i \bar{\alpha} \frac{\varepsilon_{\mathsf{u},i}^\mathsf{min} }{\sigma^4_{\mathsf{u},i}} \left| \Tilde{\textbf{h}}_{i} \textbf{f}_i \right|^2 \Tilde{\textbf{h}}_{i}^{\rm H}  \Tilde{\textbf{h}}_{i} \textbf{f}_k + \sum_{n \in \mathcal{N}} \kappa_n \frac{\varepsilon_{\mathsf{c},n}^\mathsf{min}}{\sigma^2_{\mathsf{c},n}} \left( \frac{\bar{\alpha}}{\sigma^2_{\mathsf{c},n}}  \left| \Tilde{\textbf{h}}_{n} \textbf{f}_\mathsf{c} \right|^2 \Tilde{\textbf{h}}_{n}^{\rm H}  \Tilde{\textbf{h}}_{n} \textbf{f}_k -  \Tilde{\textbf{h}}_{n}^{\rm H}  \Tilde{\textbf{h}}_{n} \textbf{f}_\mathsf{c}  \right)   
\nonumber \\
& \quad - \sum_{i \in \mathcal{G}_k}\nu_i \frac{\varepsilon_{\mathsf{u},i}^\mathsf{min}}{\sigma^2_{\mathsf{u},i}} \Tilde{\textbf{h}}_{i}^{\rm H}  \Tilde{\textbf{h}}_{i} \textbf{f}_k  + 
\left(\lambda_2 \Tilde{\textbf{H}}_\mathsf{SR}^{\rm H} \Tilde{\textbf{H}}_\mathsf{SR} + \lambda_1 \textbf{I} \right) \left(\alpha \textbf{f}_\mathsf{c} + \bar{\alpha} \textbf{f}_k\right).
\label{eq:gradient_wsr_F}\\
\nabla_{\textbf{G}} f(\textbf{F},\textbf{G},\theta ) &= \sum_{i \in \mathcal{N}}\nu_i \frac{\varepsilon_{\mathsf{u},i}^{\mathsf{min}}}{\sigma^2_{\mathsf{u},i}} \textbf{h}_i^{\rm H} \Tilde{\textbf{h}}_{i} \textbf{f}_{\mu(i)}  \textbf{f}_{\mu(i)}^{\rm H}\Bigg(\sum_{\substack{k \, {\in} \, \mathcal{K}\\ k\neq \mu(i)}}    \frac{\bar{\alpha}}{\sigma^2_{\mathsf{u},i}} \Tilde{\textbf{h}}_{i}^{\rm H}\Tilde{\textbf{h}}_{i}\textbf{f}_k \textbf{f}_k^{\rm H}  \textbf{H}_\mathsf{SR}^{\rm H} + \frac{\sigma^2}{\sigma^2_{\mathsf{u},i}} \Tilde{\textbf{h}}_{i}^{\rm H} \textbf{h}_i \textbf{G} - \textbf{H}_\mathsf{SR}^{\rm H} \Bigg)\nonumber
\\
& \quad + \sum_{i \in \mathcal{N}}  \kappa_i  \frac{\varepsilon_{\mathsf{c},i}^{\mathsf{min}}}{\sigma^2_{\mathsf{c},i}} \textbf{h}_i^{\rm H} \Tilde{\textbf{h}}_{i} \textbf{f}_\mathsf{c}  \textbf{f}_\mathsf{c}^{\rm H} \Bigg( \sum_{k \, {\in} \, \mathcal{K}} \frac{\bar{\alpha}}{\sigma^2_{\mathsf{c},i}} \Tilde{\textbf{h}}_{i}^{\rm H}  \Tilde{\textbf{h}}_{i} \textbf{f}_k \textbf{f}_k^{\rm H} \textbf{H}_\mathsf{SR}^{\rm H} + \frac{\sigma^2}{\sigma^2_{\mathsf{c},i}} \Tilde{\textbf{h}}_{i}^{\rm H} \textbf{h}_i \textbf{G} - \textbf{H}_\mathsf{SR}^{\rm H} \Bigg)\nonumber \\
&\quad + 
\lambda_2 \Tilde{\textbf{H}}_\mathsf{SR} \Bigg( \alpha \textbf{f}_\mathsf{c} \textbf{f}_\mathsf{c}^{\rm H} + \sum_{k \, {\in} \, \mathcal{K}} \bar{\alpha} \textbf{f}_k \textbf{f}_k^{\rm H} \Bigg) \textbf{H}_\mathsf{SR}^{\rm H} + \lambda_2 \sigma^2  \textbf{G}
\label{eq:gradient_wsr_G},\\
\partial_{\theta} f(\textbf{F},\textbf{G},\theta) &= -1 + \sum_{n \in \mathcal{N}} \nu_n, \hspace{3.4in} \label{eq:gradient_wsr_Rc}
\end{align}
where the detailed derivation steps for \eqref{eq:gradient_wsr_F} and \eqref{eq:gradient_wsr_G} are provided in Appendix~\ref{sec:gradient_wsr_F} and Appendix~\ref{sec:gradient_wsr_G}, respectively. 
We know from the Karush-Kuhn-Tucker (KKT) conditions that any solution to \eqref{eq:optimization_MMFSR2} requires that \eqref{eq:gradient_wsr_F}-\eqref{eq:gradient_wsr_Rc} are all equal to zero, and the following complementary slackness conditions are therefore satisfied
\begin{align}
\nu^\star_n (\theta^\star - \mathsf{R}_{\mathsf{u},n}^\star) = \kappa_n^\star (\mathsf{R}_\mathsf{c}^\mathsf{th} - \mathsf{R}_{\mathsf{c},n}^\star ) &= 0 , \label{eq:compslack_wsr_1}\\
\lambda_1^\star J_1^\star = \lambda_2^\star J_2^\star &= 0 , \label{eq:compslack_wsr_2}
\end{align}
where $^\star$ denotes the optimal value. Similarly, the gradients of \eqref{eq:lagrangian_wmmse} are given as 
\begin{align}
\nabla_{\textbf{f}_k} g(\textbf{F},\textbf{G},\eta) &= \sum_{\substack{ i \in \mathcal{N}\setminus \mathcal{G}_k}} \bar{\nu}_i v_i \bar{\alpha} \frac{(\varepsilon_{\mathsf{u},i}^\mathsf{min})^2}{\sigma^4_{\mathsf{u},i}}  |\Tilde{\textbf{h}}_i \textbf{f}_i |^2   \Tilde{\textbf{h}}_i^{\rm H} \Tilde{\textbf{h}}_i \textbf{f}_k + \left(\bar{\lambda}_2 \Tilde{\textbf{H}}_\mathsf{SR}^{\rm H} \Tilde{\textbf{H}}_\mathsf{SR} + \bar{\lambda}_1 \textbf{I} \right) \left(\alpha\textbf{f}_{\mathsf{c}} + \bar{\alpha}\textbf{f}_k\right)  
\nonumber \\
& \hspace{-0.2in} + \sum_{n \in \mathcal{N}} \bar{\kappa}_n w_n  \frac{(\varepsilon_{\mathsf{c},n}^\mathsf{min} )^2}{\sigma^2_{\mathsf{c},n}} \left( \frac{\bar{\alpha}}{\sigma^2_{\mathsf{c},n}} |\Tilde{\textbf{h}}_n \textbf{f}_{\mathsf{c}}|^2  \Tilde{\textbf{h}}_n^{\rm H}  \Tilde{\textbf{h}}_n \textbf{f}_k  - \Tilde{\textbf{h}}_n^{\rm H} \Tilde{\textbf{h}}_n \textbf{f}_{\mathsf{c}} \right) -  \sum_{ i \in \mathcal{G}_k} \bar{\nu}_i v_i \frac{(\varepsilon_{\mathsf{u},i}^\mathsf{min})^2}{\sigma^2_{\mathsf{u},i}} \Tilde{\textbf{h}}_i^{\rm H}  \Tilde{\textbf{h}}_i \textbf{f}_k 
,\label{grad_g_append} \\
\nabla_{\textbf{G}} g(\textbf{F},\textbf{G},\eta) &= 
\sum_{i \in \mathcal{N}} \bar{\nu}_i v_i  \frac{(\varepsilon_{\mathsf{u},i}^\mathsf{min})^2}{\sigma^2_{\mathsf{u},i}} \textbf{h}_i^{\rm H} \Tilde{\textbf{h}}_i \textbf{f}_{\mu(i)}  \textbf{f}_{\mu(i)}^{\rm H} \Bigg( \sum_{\substack{k \, {\in} \, \mathcal{K}\\ k\neq \mu(i)}}
\frac{\bar{\alpha}}{\sigma^2_{\mathsf{u},i}}
\Tilde{\textbf{h}}_i^{\rm H} \Tilde{\textbf{h}}_i \textbf{f}_k \textbf{f}_k^{\rm H}  \textbf{H}_\mathsf{SR}^{\rm H} + \frac{\sigma^2}{\sigma^2_{\mathsf{u},i}} \Tilde{\textbf{h}}_i^{\rm H} \textbf{h}_i \textbf{G} - \textbf{H}_\mathsf{SR}^{\rm H} \Bigg) \nonumber \\
&\quad + \sum_{i \in \mathcal{N}} \bar{\kappa}_i w_i \frac{(\varepsilon_{\mathsf{c},i}^\mathsf{min})^2}{\sigma^2_{\mathsf{c},i}}\textbf{h}_i^{\rm H} \Tilde{\textbf{h}}_i  \textbf{f}_{\mathsf{c}}\textbf{f}_{\mathsf{c}}^{\rm H} \Bigg( \sum_{k \, {\in} \, \mathcal{K}} \frac{\bar{\alpha}}{\sigma^2_{\mathsf{c},i}} \Tilde{\textbf{h}}_i^{\rm H} \Tilde{\textbf{h}}_i \textbf{f}_k \textbf{f}_k^{\rm H}  \textbf{H}_\mathsf{SR}^{\rm H} + \frac{\sigma^2}{\sigma^2_{\mathsf{c},i}} \Tilde{\textbf{h}}_i^{\rm H} \textbf{h}_i \textbf{G} -\textbf{H}_\mathsf{SR}^{\rm H}\Bigg) \nonumber \\
&\quad + \bar{\lambda}_2 \Tilde{\textbf{H}}_\mathsf{SR} \Bigg( \alpha  \textbf{f}_{\mathsf{c}} \textbf{f}_{\mathsf{c}}^{\rm H} + \sum_{k \, {\in} \, \mathcal{K}} \bar{\alpha}  \textbf{f}_k \textbf{f}_k^{\rm H} \Bigg) \textbf{H}_\mathsf{SR}^{\rm H} + \bar{\lambda}_2 \sigma^2 \textbf{G} ,
\label{grad_f_append_Gc_2} \\
\partial_{\eta} g(\textbf{F},\textbf{G},\eta) &= 1 - \sum_{n \in \mathcal{N}} \bar{\nu}_n , \label{eq:gradient_wmmse_xi}
\end{align}
where the details of \eqref{grad_g_append} and \eqref{grad_f_append_Gc_2} are given in Appendix~\ref{sec:gradient_wmmse_F} and Appendix~\ref{sec:gradient_wmmse_G}, respectively, and the corresponding complementary slackness conditions are given as
\begin{align}
\bar{\nu}_n^\star \left( \xi_{\mathsf{u},n}^{\mathsf{min} \,\star} - \eta^{\star} \right) = \bar{\kappa}^\star \left(\xi_{\mathsf{c},n}^{\mathsf{min} \,\star} - \xi_\mathsf{c}^\mathsf{th} \right) &= 0 , \label{eq:compslack_wmmse_1}\\
\bar{\lambda}_1^\star J_1^\star = \bar{\lambda}_2^\star J_2^\star &= 0 . \label{eq:compslack_wmmse_2}
\end{align}
Finally comparing \eqref{eq:gradient_wsr_F}-\eqref{eq:gradient_wsr_Rc} with  \eqref{grad_g_append}-\eqref{eq:gradient_wmmse_xi}, we observe that the gradients of the Lagrangian expression in  \eqref{eq:lagrangian_wsr} and \eqref{eq:lagrangian_wmmse} become equal if and only if \eqref{eq:optimal_weight_common} and \eqref{eq:optimal_weight_unicast} hold. \IEEEQEDhere

When we employ the optimal WMMSE weights \eqref{eq:optimal_weight_common} and \eqref{eq:optimal_weight_unicast} in the WMSE expression \eqref{eq:wmse_common} and \eqref{eq:wmse_unicast} respectively, we obtain $\xi_{\mathsf{c},n}^\mathsf{t,min} \,{=}\, 1 \,{-}\, \mathsf{R}_{\mathsf{c},n}^\mathsf{t}$ and $\xi_{\mathsf{u},n}^\mathsf{t,min} \,{=}\, 1 \,{-}\, \mathsf{R}_{\mathsf{u},n}^\mathsf{t}$, which accordingly yields $\eta \,{=}\, 1 \,{-}\, \theta$ by definition. As a result, the derivative in \eqref{eq:gradient_wsr_Rc} becomes equal to \eqref{eq:gradient_wmmse_xi}, and we have $\xi_\mathsf{c}^\mathsf{th} \,{=}\, 1 \,{-}\, \mathsf{R}_\mathsf{c}^\mathsf{th}$ as the threshold MSE for the common message. With these relations, the complementary slackness conditions in \eqref{eq:compslack_wsr_1} and \eqref{eq:compslack_wsr_2} are also equivalent to  \eqref{eq:compslack_wmmse_1} and \eqref{eq:compslack_wmmse_2}, respectively. As a general interpretation, note also that we have $\mathsf{R}_{\mathsf{c},n}^\mathsf{t} \,{=}\, \mathsf{R}_\mathsf{c}^\mathsf{th}$ and  $\mathsf{R}_{\mathsf{u},n}^\mathsf{t} \,{=}\, \theta$ for $\kappa_n^\mathsf{t} \,{>}\, 0$ and $\nu_n^\mathsf{t} \,{>}\, 0$, and equivalently  $\xi_{\mathsf{c},n}^\mathsf{t,min} \,{=}\, \xi_\mathsf{c}^\mathsf{th}$ and $\xi_{\mathsf{u},n}^\mathsf{t,min} \,{=}\, \eta $ for $\bar{\kappa}_n^\mathsf{t} \,{>}\, 0$ and $\bar{\nu}_n^\mathsf{t} \,{>}\, 0$. Similarly, whenever we have $\kappa_n^\mathsf{t} \,{=}\, 0$ ($\nu_n^\mathsf{t} \,{=}\, 0$), we also have $\bar{\kappa}_n^\mathsf{t} \,{=}\, 0$ ($\bar{\nu}_n^\mathsf{t} \,{=}\, 0$).
\color{black}

\section{} \label{sec:gradient_wsr_F}
In this section, we derive $\nabla_{\textbf{f}_k^\mathsf{\,t}}f\left(\textbf{F}^\mathsf{t},\textbf{G}^\mathsf{t},\eta_\mathsf{c} \right)$ for SC scheme (i.e., $\mathsf{\,t} \,{=}\, \mathsf{SC}$). For notational convenience, we drop the superscript $\mathsf{\,t}$ for the rest of the derivation. The Lagrangian objective function in \eqref{eq:lagrangian_wsr} is split into various terms, as below, to compute the gradient term-by-term basis.
\begin{align}
f\left(\textbf{F}^\mathsf{t},\textbf{G}^\mathsf{t},\theta \right) &= -  \theta + \underbrace{ \sum_{n \in \mathcal{N}} \nu_n^\mathsf{t}\left(\theta - \mathsf{R}_{\mathsf{u},n}^\mathsf{t} \right)}_\text{A} + \underbrace{ \sum_{n \in \mathcal{N}} \kappa_n^\mathsf{t} \left(\mathsf{R}_\mathsf{c}^\mathsf{th} - \mathsf{R}_{\mathsf{c},n}^\mathsf{t}\right) }_\text{B} + \underbrace{ \lambda_1^\mathsf{t} J_1^\mathsf{t}}_\text{C} + \underbrace{ \lambda_2^\mathsf{t} J_2^\mathsf{t}}_\text{D}. \label{append_lagrangian_SR}
\end{align}
We begin with calculating the gradient of A in (\ref{append_lagrangian_SR}). First, we need $\nabla_{\textbf{f}_k} \mathsf{R}_{\mathsf{u},i}$ for both $\mu(i)=k$ (i.e., $i \in \mathcal{G}_k$) and $\mu(i) \neq k$, (i.e., $i \in \mathcal{N} \setminus \mathcal{G}_k$). Starting with the case $\mu(i)=k$, and using (\ref{eq:mmse_unicast}) and (\ref{eq:rate_mmse_unicast}), we obtain $\nabla_{\textbf{f}_k}\mathsf{R}_{\mathsf{u},i}= (\nabla_{\textbf{f}_k}\varepsilon_{\mathsf{u},i}^\mathsf{min^{-1}}) \varepsilon_{\mathsf{u},i}^\mathsf{min}$. Recalling that the noise variance $\frac{1}{\sigma^2_{\mathsf{u},i}}$ is independent from $\textbf{f}_k$, and $\nabla_{\textbf{X}}(\textbf{X}^{\rm H}\textbf{A}\textbf{X})= \textbf{A}\textbf{X}$ \cite{moon_stirling_math}, we obtain 
\begin{align}
\nabla_{[\textbf{f}_k]_{m}} \varepsilon_{\mathsf{u},i}^\mathsf{min^{-1}} &= \textbf{e}_m^{\rm H} \textbf{H}_\mathsf{SR}^{\rm H} \textbf{G}^{\rm H} \textbf{h}_i^{\rm H} \frac{1}{\sigma^2_{\mathsf{u},i}} \textbf{h}_i \textbf{G} \textbf{H}_\mathsf{SR} \textbf{f}_k , \label{grad_cov_E_eps_k_mn}
\end{align}
where $\textbf{e}_m$ is $M{\times}1$ unity column vector with $1$ at the $m$-th element and zeros elsewhere. Since $[\nabla_{\textbf{f}_k}\mathsf{R}_{\mathsf{u},i}]_{m} = \nabla_{[\textbf{f}_k]_{m}}\mathsf{R}_{\mathsf{u},i}= \textbf{e}_m^{\rm H} \textbf{H}_\mathsf{SR}^{\rm H} \textbf{G}^{\rm H} \textbf{h}_i^{\rm H} \frac{1}{\sigma^2_{\mathsf{u},i}} \textbf{h}_i \textbf{G} \textbf{H}_\mathsf{SR} \textbf{f}_k \varepsilon_{\mathsf{u},i}^\mathsf{min} $, we have
\begin{align}
\nabla_{\textbf{f}_k} \mathsf{R}_{\mathsf{u},i} &= \textbf{H}_\mathsf{SR}^{\rm H} \textbf{G}^{\rm H} \textbf{h}_i^{\rm H} \frac{1}{\sigma^2_{\mathsf{u},i}} \textbf{h}_i \textbf{G} \textbf{H}_\mathsf{SR} \textbf{f}_k \varepsilon_{\mathsf{u},i}^\mathsf{min} \label{grad_R_k}.
\end{align}We then consider $\mu(i) \neq k$ for which $\nabla_{\textbf{f}_k}\mathsf{R}_{\mathsf{u},i}$ is computed as follows
\begin{align}
\nabla_{[\textbf{f}_k]_{m}}\mathsf{R}_{\mathsf{u},i}
&= \textbf{f}_i^{\rm H} \textbf{H}_\mathsf{SR}^{\rm H} \textbf{G}^{\rm H} \textbf{h}_i^{\rm H} \nabla_{[\textbf{f}_k]_{m}}\left(\frac{1}{\sigma^2_{\mathsf{u},i}}\right) \textbf{h}_i \textbf{G} \textbf{H}_\mathsf{SR} \textbf{f}_i \varepsilon_{\mathsf{u},i}^\mathsf{min}.\label{grad_cov_E_eps_i}
\end{align}
Using $\nabla_{\textbf{X}} (\textbf{X}^{-1}) = -\textbf{X}^{-1}\nabla (\textbf{X}) \textbf{X}^{-1}$ \cite{matrix_cookbook}, we can write
\begin{align}
\nabla_{[\textbf{f}_k]_{m}} \left(\sigma^{{-}2}_{\mathsf{u},i} \right) &= -\frac{1}{\sigma^2_{\mathsf{u},i}}\nabla_{[\textbf{f}_k]_{m}}(\sigma^2_{\mathsf{u},i})\frac{1}{\sigma^2_{\mathsf{u},i}}, \label{grad_pkm_invrn2i}
\end{align}
where
\begin{align}
\nabla_{[\textbf{f}_k]_{m}}(\sigma^2_{\mathsf{u},i})&=  \bar{\alpha} \textbf{h}_i \textbf{G} \textbf{H}_\mathsf{SR} \textbf{f}_k  \textbf{e}_m^{\rm H} \textbf{H}_\mathsf{SR}^{\rm H} \textbf{G}^{\rm H} \textbf{h}_i^{\rm H}.\label{grad_pkm_rn2i}
\end{align}
By combining (\ref{grad_cov_E_eps_i}), (\ref{grad_pkm_invrn2i}) and (\ref{grad_pkm_rn2i}), we yield
\begin{align}
 \nabla_{[\textbf{f}_k]_{m}}\mathsf{R}_{\mathsf{u},i}&= -\bar{\alpha}\textbf{e}_m^{\rm H} \textbf{H}_\mathsf{SR}^{\rm H} \textbf{G}^{\rm H} \textbf{h}_i^{\rm H}\frac{1}{\sigma^2_{\mathsf{u},i}} \textbf{h}_i \textbf{G} \textbf{H}_\mathsf{SR} \textbf{f}_i {\varepsilon}_{\mathsf{u},i}^\mathsf{min}  \textbf{f}_i^{\rm H} \textbf{H}_\mathsf{SR}^{\rm H} \textbf{G}^{\rm H} \textbf{h}_i^{\rm H} \frac{1}{\sigma^2_{\mathsf{u},i}} \textbf{h}_i \textbf{G} \textbf{H}_\mathsf{SR} \textbf{f}_k, \label{grad_cov_R_eps_i_mn}
\end{align}
which can be represented considering all the entries of the gradient as follows
\begin{align}
\nabla_{\textbf{f}_k}\mathsf{R}_{\mathsf{u},i} &= -\bar{\alpha}\textbf{H}_\mathsf{SR}^{\rm H} \textbf{G}^{\rm H} \textbf{h}_i^{\rm H}\frac{1}{\sigma^2_{\mathsf{u},i}} \textbf{h}_i \textbf{G} \textbf{H}_\mathsf{SR} \textbf{f}_i {\varepsilon}_{\mathsf{u},i}^\mathsf{min}  \textbf{f}_i^{\rm H} \textbf{H}_\mathsf{SR}^{\rm H} \textbf{G}^{\rm H} \textbf{h}_i^{\rm H} \frac{1}{\sigma^2_{\mathsf{u},i}} \textbf{h}_i \textbf{G} \textbf{H}_\mathsf{SR} \textbf{f}_k. \label{grad_R_i_last}
\end{align}
Finally, combining (\ref{grad_R_k}) and (\ref{grad_R_i_last}), the gradient of A is given as follows
\begin{align}
\nabla_{\textbf{f}_k} \mathrm{A} &= \sum_{i \in \mathcal{N}\setminus \mathcal{G}_k} \nu_i \bar{\alpha} \frac{\varepsilon_{\mathsf{u},i}^\mathsf{min} }{\sigma^4_{\mathsf{u},i}} \textbf{H}_\mathsf{SR}^{\rm H} \textbf{G}^{\rm H} \textbf{h}_i^{\rm H} \textbf{h}_i \textbf{G} \textbf{H}_\mathsf{SR} \textbf{f}_i  \textbf{f}_i^{\rm H} \textbf{H}_\mathsf{SR}^{\rm H} \textbf{G}^{\rm H} \textbf{h}_i^{\rm H}  \textbf{h}_i \textbf{G} \textbf{H}_\mathsf{SR} \textbf{f}_k\nonumber \\
& \qquad - \sum_{i \in \mathcal{G}_k}\nu_i \frac{\varepsilon_{\mathsf{u},i}^\mathsf{min}}{\sigma^2_{\mathsf{u},i}} \textbf{H}_\mathsf{SR}^{\rm H} \textbf{G}^{\rm H} \textbf{h}_i^{\rm H}  \textbf{h}_i \textbf{G} \textbf{H}_\mathsf{SR} \textbf{f}_k  \label{grad_B_last}.
\end{align}Secondly, we compute the gradient of B in (\ref{append_lagrangian_SR}). We first note that $\nabla_{\textbf{f}_k} \mathsf{R}_{\mathsf{c},n} = (\nabla \varepsilon_{\mathsf{c},n}^\mathsf{min^{-1}}) \varepsilon_{\mathsf{c},n}^\mathsf{min}$. The gradient $\nabla_{[\textbf{f}_k]_{m}} \varepsilon_{\mathsf{c},n}^\mathsf{min^{-1}}$ is therefore calculated by applying the chain rule to $\nabla_{[\textbf{f}_k]_{m}} \varepsilon_{\mathsf{c},n}^\mathsf{min^{-1}}$, which is represented as follows
\begin{align}
\nabla_{[\textbf{f}_k]_{m}} \varepsilon_{\mathsf{c},n}^\mathsf{min^{-1}} &= \nabla_{[\textbf{f}_k]_{m}} \left ({\frac{1}{\alpha} } +  \textbf{f}_{\mathsf{c}}^{\rm H}\textbf{H}_\mathsf{SR}^{\rm H} \textbf{G}^{\rm H} \textbf{h}_n^{\rm H} \frac{1}{\sigma^2_{\mathsf{c},n}}\textbf{h}_n \textbf{G} \textbf{H}_\mathsf{SR} \textbf{f}_{\mathsf{c}}\right),  \\
&=  \textbf{f}_{\mathsf{c}}^{\rm H}\textbf{H}_\mathsf{SR}^{\rm H} \textbf{G}^{\rm H} \textbf{h}_n^{\rm H} \frac{\partial }{[\partial\textbf{f}_k^\ast]_{m}}\left(\frac{1}{\sigma^2_{\mathsf{c},n}}\right)\textbf{h}_n \textbf{G} \textbf{H}_\mathsf{SR} \textbf{f}_{\mathsf{c}} +  \frac{\partial (\textbf{f}_{\mathsf{c}}^{\rm H}\textbf{H}_\mathsf{SR}^{\rm H} \textbf{G}^{\rm H} \textbf{h}_n^{\rm H})}{[\partial \textbf{f}_k^\ast]_{m}}\frac{1}{\sigma^2_{\mathsf{c},n}}\textbf{h}_n \textbf{G} \textbf{H}_\mathsf{SR} \textbf{f}_{\mathsf{c}} \nonumber \\
&\quad + \textbf{f}_{\mathsf{c}}^{\rm H}\textbf{H}_\mathsf{SR}^{\rm H} \textbf{G}^{\rm H} \textbf{h}_n^{\rm H} \frac{1}{\sigma^2_{\mathsf{c},n}} \frac{\partial (\textbf{h}_n \textbf{G} \textbf{H}_\mathsf{SR} \textbf{f}_{\mathsf{c}})}{[\partial\textbf{f}_k^\ast]_{m}} ,  \\
&= -\textbf{f}_{\mathsf{c}}^{\rm H}\textbf{H}_\mathsf{SR}^{\rm H} \textbf{G}^{\rm H} \textbf{h}_n^{\rm H} \frac{1}{\sigma^2_{\mathsf{c},n}} \frac{\partial (\sigma^2_{\mathsf{c},n})}{[\partial\textbf{f}_k^\ast]_{m}} \frac{1}{\sigma^2_{\mathsf{c},n}}\textbf{h}_n \textbf{G} \textbf{H}_\mathsf{SR} \textbf{f}_{\mathsf{c}}  + \textbf{e}_m^{\rm H} \textbf{H}_\mathsf{SR}^{\rm H} \textbf{G}^{\rm H} \textbf{h}_n^{\rm H} \frac{1}{\sigma^2_{\mathsf{c},n}}\textbf{h}_n \textbf{G} \textbf{H}_\mathsf{SR} \textbf{f}_{\mathsf{c}} .
\end{align}
Noting that $\frac{\partial\textbf{f}_k}{\partial \textbf{f}_k^\ast} = 0$, and $\nabla_{\textbf{X}}(\textbf{X}\textbf{A}\textbf{X}^{\rm H})= \textbf{X}\textbf{A}$\cite{moon_stirling_math}, we obtain
\begin{align}
\nabla_{[\textbf{f}_k]_m} \sigma^2_{\mathsf{c},n} &= \bar{\alpha} \textbf{h}_n \textbf{G} \textbf{H}_\mathsf{SR} \textbf{f}_k  \textbf{e}_m^{\rm H} \textbf{H}_\mathsf{SR}^{\rm H} \textbf{G}^{\rm H} \textbf{h}_n^{\rm H}.
\end{align}
We then have
\begin{align}
\nabla_{[\textbf{f}_k]_{m}}\varepsilon_{\mathsf{c},n}^\mathsf{min^{-1}} &= -\textbf{f}_{\mathsf{c}}^{\rm H}\textbf{H}_\mathsf{SR}^{\rm H} \textbf{G}^{\rm H} \textbf{h}_n^{\rm H} \frac{1}{\sigma^2_{\mathsf{c},n}} \bar{\alpha} \textbf{h}_n \textbf{G} \textbf{H}_\mathsf{SR} \textbf{f}_k  \textbf{e}_m^{\rm H} \textbf{H}_\mathsf{SR}^{\rm H} \textbf{G}^{\rm H} \textbf{h}_n^{\rm H} \frac{1}{\sigma^2_{\mathsf{c},n}}\textbf{h}_n \textbf{G} \textbf{H}_\mathsf{SR} \textbf{f}_{\mathsf{c}}  \nonumber \\
&\quad + \textbf{e}_m^{\rm H} \textbf{H}_\mathsf{SR}^{\rm H} \textbf{G}^{\rm H} \textbf{h}_n^{\rm H} \frac{1}{\sigma^2_{\mathsf{c},n}}\textbf{h}_n \textbf{G} \textbf{H}_\mathsf{SR} \textbf{f}_{\mathsf{c}}, \label{append_invEck}\\
\nabla_{[\textbf{f}_k]_{m}}\mathsf{R}_{\mathsf{c},n} &= - \bar{\alpha} \textbf{e}_m^{\rm H} \textbf{H}_\mathsf{SR}^{\rm H} \textbf{G}^{\rm H} \textbf{h}_n^{\rm H} \frac{1}{\sigma^2_{\mathsf{c},n}}\textbf{h}_n \textbf{G} \textbf{H}_\mathsf{SR} \textbf{f}_{\mathsf{c}} \varepsilon_{\mathsf{c},n}^\mathsf{min} \textbf{f}_{\mathsf{c}}^{\rm H}\textbf{H}_\mathsf{SR}^{\rm H} \textbf{G}^{\rm H} \textbf{h}_n^{\rm H} \frac{1}{\sigma^2_{\mathsf{c},n}} \textbf{h}_n \textbf{G} \textbf{H}_\mathsf{SR} \textbf{f}_k   \nonumber \\
&\quad + \textbf{e}_m^{\rm H} \textbf{H}_\mathsf{SR}^{\rm H} \textbf{G}^{\rm H} \textbf{h}_n^{\rm H} \frac{1}{\sigma^2_{\mathsf{c},n}}\textbf{h}_n \textbf{G} \textbf{H}_\mathsf{SR} \textbf{f}_{\mathsf{c}} \varepsilon_{\mathsf{c},n}^\mathsf{min}.
 \label{grad_R_eps_cl_mn_part1}
\end{align}Finally, we obtain
\begin{align}
\nabla_{\textbf{f}_k}\mathsf{R}_{\mathsf{c},n} = &- \bar{\alpha} \textbf{H}_\mathsf{SR}^{\rm H} \textbf{G}^{\rm H} \textbf{h}_n^{\rm H} \frac{1}{\sigma^2_{\mathsf{c},n}}\textbf{h}_n \textbf{G} \textbf{H}_\mathsf{SR} \textbf{f}_{\mathsf{c}} \varepsilon_{\mathsf{c},n}^\mathsf{min} \textbf{f}_{\mathsf{c}}^{\rm H}\textbf{H}_\mathsf{SR}^{\rm H} \textbf{G}^{\rm H} \textbf{h}_n^{\rm H} \frac{1}{\sigma^2_{\mathsf{c},n}} \textbf{h}_n \textbf{G} \textbf{H}_\mathsf{SR} \textbf{f}_k   \nonumber \\
&\quad + \textbf{H}_\mathsf{SR}^{\rm H} \textbf{G}^{\rm H} \textbf{h}_n^{\rm H} \frac{1}{\sigma^2_{\mathsf{c},n}}\textbf{h}_n \textbf{G} \textbf{H}_\mathsf{SR} \textbf{f}_{\mathsf{c}} \varepsilon_{\mathsf{c},n}^\mathsf{min}, \label{grad_R_eps_cl_last}
\end{align}and the gradient of B becomes
\begin{align}
\nabla_{\textbf{f}_k} \mathrm{B}
= &\sum_{n \in \mathcal{N}} \kappa_n \bar{\alpha} \frac{\varepsilon_{\mathsf{c},n}^\mathsf{min}}{\sigma^4_{\mathsf{c},n}} \textbf{H}_\mathsf{SR}^{\rm H} \textbf{G}^{\rm H} \textbf{h}_n^{\rm H} \textbf{h}_n \textbf{G} \textbf{H}_\mathsf{SR} \textbf{f}_{\mathsf{c}}  \textbf{f}_{\mathsf{c}}^{\rm H}\textbf{H}_\mathsf{SR}^{\rm H} \textbf{G}^{\rm H} \textbf{h}_n^{\rm H} \textbf{h}_n \textbf{G} \textbf{H}_\mathsf{SR} \textbf{f}_k  \nonumber \\
&- \sum_{n \in \mathcal{N}} \kappa_n \frac{\varepsilon_{\mathsf{c},n}^\mathsf{min}}{\sigma^2_{\mathsf{c},n}} \textbf{H}_\mathsf{SR}^{\rm H} \textbf{G}^{\rm H} \textbf{h}_n^{\rm H} \textbf{h}_n \textbf{G} \textbf{H}_\mathsf{SR} \textbf{f}_{\mathsf{c}}  .
\label{grad_A_last}
\end{align}
Finally, we compute the gradients of C and D in (\ref{append_lagrangian_SR}). We use the identity $\nabla_{\textbf{f}_k} \Tr (\textbf{f}_{\mathsf{c}} \textbf{f}_{\mathsf{c}}^{\rm H}) = \textbf{f}_{\mathsf{c}}$, which obtains the following expressions
\begin{align}
\nabla_{\textbf{f}_k} \mathrm{C}
&= \lambda_1 \left(\alpha\textbf{f}_{\mathsf{c}} + \bar{\alpha}\textbf{f}_k\right),
\label{grad_C_last}\\
\nabla_{\textbf{f}_k} \mathrm{D}
&= \lambda_2 \left(\alpha \textbf{H}_\mathsf{SR}^{\rm H} \textbf{G}^{\rm H}\textbf{G} \textbf{H}_\mathsf{SR} \textbf{f}_{\mathsf{c}} + \bar{\alpha}\textbf{H}_\mathsf{SR}^{\rm H} \textbf{G}^{\rm H}\textbf{G} \textbf{H}_\mathsf{SR} \textbf{f}_k\right).
\label{grad_D_last}
\end{align}
We finally combine (\ref{grad_B_last}), (\ref{grad_A_last}), (\ref{grad_C_last}) and (\ref{grad_D_last}) to yield the desired gradient in \eqref{eq:gradient_wsr_F}.
\section{}\label{sec:gradient_wsr_G}
In this section, we derive $\nabla_{\textbf{G}^\mathsf{t}}f\left(\textbf{F}^\mathsf{t},\textbf{G}^\mathsf{t}, \theta \right)$ for $\mathsf{\,t} = \mathsf{SC}$. For notational convenience, we drop the superscript $\mathsf{\,t}$. We first calculate the gradient of A in (\ref{append_lagrangian_SR}). We have $\nabla_{\textbf{G}} \mathsf{R}_{\mathsf{u},i}= \left(\nabla_{\textbf{G}}\varepsilon_{\mathsf{u},i}^{\mathsf{min}^{-1}}\right) \varepsilon_{\mathsf{u},i}^\mathsf{min}$.
 The gradient is a matrix with the $[m,n]$-th element defined as $\nabla_{[\textbf{G}]_{mn}}$. Note that the noise variance $\frac{1}{\sigma^2_{\mathsf{u},i}}$ is not independent from $\textbf{G}$. Thus, the gradient $\nabla_{[\textbf{G}]_{mn}} \varepsilon_{\mathsf{u},i}^{\mathsf{min}^{-1}}$ is calculated by applying the chain rule on $\nabla_{[\textbf{G}]_{mn}} \varepsilon_{\mathsf{u},i}^{\mathsf{min}^{-1}}$.
\begin{align}
\nabla_{[\textbf{G}]_{mn}} \varepsilon_{\mathsf{u},i}^{\mathsf{min}^{-1}} &= \nabla_{[\textbf{G}]_{mn}} \left ({\frac{1}{\bar{\alpha}} } +  \textbf{f}_{\mu(i)}^{\rm H}\textbf{H}_\mathsf{SR}^{\rm H} \textbf{G}^{\rm H} \textbf{h}_i^{\rm H} \frac{1}{\sigma^2_{\mathsf{u},i}}\textbf{h}_i \textbf{G} \textbf{H}_\mathsf{SR} \textbf{f}_{\mu(i)}\right) \nonumber \\
&=  \textbf{f}_{\mu(i)}^{\rm H}\textbf{H}_\mathsf{SR}^{\rm H} \textbf{G}^{\rm H} \textbf{h}_i^{\rm H} \frac{\partial}{[\partial\textbf{G}^\ast]_{mn}} \left(\frac{1}{\sigma^2_{\mathsf{u},i}}\right)\textbf{h}_i \textbf{G} \textbf{H}_\mathsf{SR} \textbf{f}_{\mu(i)} +  \frac{\partial (\textbf{f}_{\mu(i)}^{\rm H}\textbf{H}_\mathsf{SR}^{\rm H} \textbf{G}^{\rm H} \textbf{h}_i^{\rm H})}{[\partial \textbf{G}^\ast]_{mn}}\frac{1}{\sigma^2_{\mathsf{u},i}}\textbf{h}_i \textbf{G} \textbf{H}_\mathsf{SR} \textbf{f}_{\mu(i)} \nonumber \\
&\quad + \textbf{f}_{\mu(i)}^{\rm H}\textbf{H}_\mathsf{SR}^{\rm H} \textbf{G}^{\rm H} \textbf{h}_i^{\rm H} \frac{1}{\sigma^2_{\mathsf{u},i}} \frac{\partial (\textbf{h}_i \textbf{G} \textbf{H}_\mathsf{SR} \textbf{f}_{\mu(i)})}{[\partial\textbf{G}^\ast]_{mn}} \nonumber  \\
&= -\textbf{f}_{\mu(i)}^{\rm H}\textbf{H}_\mathsf{SR}^{\rm H} \left( \textbf{G}^{\rm H} \textbf{h}_i^{\rm H} \frac{1}{\sigma^2_{\mathsf{u},i}} \frac{\partial (\sigma^2_{\mathsf{u},i})}{[\partial\textbf{G}^\ast]_{mn}} \frac{1}{\sigma^2_{\mathsf{u},i}}\textbf{h}_i \textbf{G}   -    \textbf{e}_n \textbf{e}_m^{\rm H} \textbf{h}_i^{\rm H} \frac{1}{\sigma^2_{\mathsf{u},i}}\textbf{h}_i \textbf{G}\right) \textbf{H}_\mathsf{SR} \textbf{f}_{\mu(i)} .
\end{align}Now we continue to compute
\begin{align}
\nabla_{[\textbf{G}]_{mn}} \sigma^2_{\mathsf{u},i} &= \sum_{\substack{k \, {\in} \, \mathcal{K}\\ k\neq \mu(i)}} \bar{\alpha} \textbf{h}_i \textbf{G} \textbf{H}_\mathsf{SR} \textbf{f}_k \textbf{f}_k^{\rm H}  \textbf{H}_\mathsf{SR}^{\rm H} \textbf{e}_n \textbf{e}_m^{\rm H} \textbf{h}_i^{\rm H} + \sigma^2\textbf{h}_i \textbf{G} \textbf{e}_n \textbf{e}_m^{\rm H} \textbf{h}_i^{\rm H},
\end{align}as $\nabla_{\textbf{X}}(\textbf{X}\textbf{A}\textbf{X}^{\rm H})= \textbf{X}\textbf{A}$\cite{moon_stirling_math}. Then, we have
\begin{align}
\nabla_{[\textbf{G}]_{mn}}\varepsilon_{\mathsf{u},i}^{\mathsf{min}^{-1}} &= -  \sum_{\substack{k \, {\in} \, \mathcal{K}\\ k\neq \mu(i)}} \textbf{f}_{\mu(i)}^{\rm H}\textbf{H}_\mathsf{SR}^{\rm H} \textbf{G}^{\rm H} \textbf{h}_i^{\rm H} \frac{1}{\sigma^2_{\mathsf{u},i}} \bar{\alpha} \textbf{h}_i \textbf{G} \textbf{H}_\mathsf{SR} \textbf{f}_k \textbf{f}_k^{\rm H}  \textbf{H}_\mathsf{SR}^{\rm H} \textbf{e}_n \textbf{e}_m^{\rm H} \textbf{h}_i^{\rm H} \frac{1}{\sigma^2_{\mathsf{u},i}}\textbf{h}_i \textbf{G} \textbf{H}_\mathsf{SR} \textbf{f}_{\mu(i)} \nonumber \\
& \quad - \textbf{f}_{\mu(i)}^{\rm H}\textbf{H}_\mathsf{SR}^{\rm H}\left( \textbf{G}^{\rm H} \textbf{h}_i^{\rm H} \frac{1}{\sigma^4_{\mathsf{u},i}} \sigma^2\textbf{h}_i \textbf{G} \textbf{e}_n \textbf{e}_m^{\rm H} \textbf{h}_i^{\rm H} \textbf{h}_i \textbf{G}  -  \textbf{e}_n \textbf{e}_m^{\rm H} \textbf{h}_i^{\rm H} \frac{1}{\sigma^2_{\mathsf{u},i}}\textbf{h}_i \textbf{G}\right) \textbf{H}_\mathsf{SR} \textbf{f}_{\mu(i)}, \label{append_invEck_G}\\
\nabla_{[\textbf{G}]_{mn}}\mathsf{R}_{\mathsf{u},i} &= -  \sum_{k \, {\in} \, \mathcal{K}, k\neq \mu(i)}  \textbf{e}_m^{\rm H} \bar{\alpha} \frac{\varepsilon_{\mathsf{u},i}^{\mathsf{min}} }{\sigma^4_{\mathsf{u},i}} \textbf{h}_i^{\rm H} \textbf{h}_i \textbf{G} \textbf{H}_\mathsf{SR} \textbf{f}_{\mu(i)} \textbf{f}_{\mu(i)}^{\rm H}\textbf{H}_\mathsf{SR}^{\rm H} \textbf{G}^{\rm H} \textbf{h}_i^{\rm H}   \textbf{h}_i \textbf{G} \textbf{H}_\mathsf{SR} \textbf{f}_k \textbf{f}_k^{\rm H}  \textbf{H}_\mathsf{SR}^{\rm H} \textbf{e}_n \nonumber \\
& \quad - \textbf{e}_m^{\rm H} \frac{\varepsilon_{\mathsf{u},i}^{\mathsf{min}}}{\sigma^2_{\mathsf{u},i}} \textbf{h}_i^{\rm H} \textbf{h}_i \textbf{G} \textbf{H}_\mathsf{SR} \textbf{f}_{\mu(i)} \textbf{f}_{\mu(i)}^{\rm H}\textbf{H}_\mathsf{SR}^{\rm H} \left(\frac{\sigma^2 }{\sigma^2_{\mathsf{u},i}}  \textbf{G}^{\rm H} \textbf{h}_i^{\rm H} \textbf{h}_i \textbf{G}  -  \textbf{I} \right)\textbf{e}_n .
 \label{grad_R_eps_cl_mn_part1_G}
\end{align}Finally, we obtain
\begin{align}
\nabla_{\textbf{G}}\mathsf{R}_{\mathsf{u},i} &= -  \sum_{\substack{k \, {\in} \, \mathcal{K}\\ k\neq \mu(i)}}   \bar{\alpha} \frac{\varepsilon_{\mathsf{u},i}^{\mathsf{min}}}{\sigma^4_{\mathsf{u},i}} \textbf{h}_i^{\rm H} \textbf{h}_i \textbf{G} \textbf{H}_\mathsf{SR} \textbf{f}_{\mu(i)} \textbf{f}_{\mu(i)}^{\rm H}\textbf{H}_\mathsf{SR}^{\rm H} \textbf{G}^{\rm H} \textbf{h}_i^{\rm H}   \textbf{h}_i \textbf{G} \textbf{H}_\mathsf{SR} \textbf{f}_k \textbf{f}_k^{\rm H}  \textbf{H}_\mathsf{SR}^{\rm H}  \nonumber \\
& \quad -  \frac{\sigma^2 \varepsilon_{\mathsf{u},i}^{\mathsf{min}}}{\sigma^4_{\mathsf{u},i}} \textbf{h}_i^{\rm H} \textbf{h}_i \textbf{G} \textbf{H}_\mathsf{SR} \textbf{f}_{\mu(i)} \textbf{f}_{\mu(i)}^{\rm H}\textbf{H}_\mathsf{SR}^{\rm H} \textbf{G}^{\rm H} \textbf{h}_i^{\rm H} \textbf{h}_i \textbf{G}  +  \frac{\varepsilon_{\mathsf{u},i}^{\mathsf{min}}}{\sigma^2_{\mathsf{u},i}} \textbf{h}_i^{\rm H} \textbf{h}_i \textbf{G} \textbf{H}_\mathsf{SR} \textbf{f}_{\mu(i)}  \textbf{f}_{\mu(i)}^{\rm H}\textbf{H}_\mathsf{SR}^{\rm H}, \label{grad_R_eps_cl_last_G}
\end{align}and the gradient of A becomes
\begin{align}
\nabla_{\textbf{G}} \mathrm{A}
& = \sum_{i \in \mathcal{N}} \sum_{\substack{k \, {\in} \, \mathcal{K}\\ k\neq \mu(i)}}  \nu_i \bar{\alpha} \frac{\varepsilon_{\mathsf{u},i}^{\mathsf{min}}}{\sigma^4_{\mathsf{u},i}} \textbf{h}_i^{\rm H} \textbf{h}_i \textbf{G} \textbf{H}_\mathsf{SR} \textbf{f}_{\mu(i)} \textbf{f}_{\mu(i)}^{\rm H}\textbf{H}_\mathsf{SR}^{\rm H} \textbf{G}^{\rm H} \textbf{h}_i^{\rm H}   \textbf{h}_i \textbf{G} \textbf{H}_\mathsf{SR} \textbf{f}_k \textbf{f}_k^{\rm H}  \textbf{H}_\mathsf{SR}^{\rm H}  \nonumber \\
& \quad + \sum_{i \in \mathcal{N}}  \nu_i \frac{\varepsilon_{\mathsf{u},i}^{\mathsf{min}}}{\sigma^2_{\mathsf{u},i}} \textbf{h}_i^{\rm H} \textbf{h}_i \textbf{G} \textbf{H}_\mathsf{SR} \textbf{f}_{\mu(i)}  \textbf{f}_{\mu(i)}^{\rm H}\textbf{H}_\mathsf{SR}^{\rm H} \left( \frac{\sigma^2 }{\sigma^2_{\mathsf{u},i}} \textbf{G}^{\rm H} \textbf{h}_i^{\rm H} \textbf{h}_i \textbf{G}  - \textbf{I} \right).
\label{grad_A_last_Gc}
\end{align}
Now, we calculate the gradient of B in (\ref{append_lagrangian_SR}). We have $\nabla_{\textbf{G}} \mathsf{R}_{\mathsf{c},i}= \left(\nabla_{\textbf{G}}\varepsilon_{\mathsf{c},i}^{\mathsf{min}^{-1}}\right) \varepsilon_{\mathsf{c},i}^\mathsf{min}$. Note that the noise variance $\frac{1}{\sigma^2_{\mathsf{c},i}}$ is not independent from $\textbf{G}$. Thus, the gradient $\nabla_{[\textbf{G}]_{mn}}\varepsilon_{\mathsf{c},i}^{\mathsf{min}^{-1}}$ is calculated by applying the chain rule on $\nabla_{[\textbf{G}]_{mn}} \varepsilon_{\mathsf{c},i}^{\mathsf{min}^{-1}}$
\begin{align}
\nabla_{[\textbf{G}]_{mn}} \varepsilon_{\mathsf{c},i}^{\mathsf{min}^{-1}} &= \nabla_{[\textbf{G}]_{mn}} \left({\frac{1}{\alpha}} +  \textbf{f}_{\mathsf{c}}^{\rm H}\textbf{H}_\mathsf{SR}^{\rm H} \textbf{G}^{\rm H} \textbf{h}_i^{\rm H} \frac{1}{\sigma^2_{\mathsf{c},i}}\textbf{h}_i \textbf{G} \textbf{H}_\mathsf{SR} \textbf{f}_{\mathsf{c}}\right) \nonumber \\
&=  \textbf{f}_{\mathsf{c}}^{\rm H}\textbf{H}_\mathsf{SR}^{\rm H} \textbf{G}^{\rm H} \textbf{h}_i^{\rm H} \frac{\partial }{[\partial\textbf{G}^\ast]_{mn}}\left(\frac{1}{\sigma^2_{\mathsf{c},i}}\right)\textbf{h}_i \textbf{G} \textbf{H}_\mathsf{SR} \textbf{f}_{\mathsf{c}} +  \frac{\partial (\textbf{f}_{\mathsf{c}}^{\rm H}\textbf{H}_\mathsf{SR}^{\rm H} \textbf{G}^{\rm H} \textbf{h}_i^{\rm H})}{[\partial \textbf{G}^\ast]_{mn}}\frac{1}{\sigma^2_{\mathsf{c},i}}\textbf{h}_i \textbf{G} \textbf{H}_\mathsf{SR} \textbf{f}_{\mathsf{c}} \nonumber \\
&\quad + \textbf{f}_{\mathsf{c}}^{\rm H}\textbf{H}_\mathsf{SR}^{\rm H} \textbf{G}^{\rm H} \textbf{h}_i^{\rm H} \frac{1}{\sigma^2_{\mathsf{c},i}} \frac{\partial (\textbf{h}_i \textbf{G} \textbf{H}_\mathsf{SR} \textbf{f}_{\mathsf{c}})}{[\partial\textbf{G}^\ast]_{mn}} \nonumber  \\
&= -\textbf{f}_{\mathsf{c}}^{\rm H}\textbf{H}_\mathsf{SR}^{\rm H} \textbf{G}^{\rm H} \textbf{h}_i^{\rm H} \frac{1}{\sigma^2_{\mathsf{c},i}} \frac{\partial (\sigma^2_{\mathsf{c},i})}{[\partial\textbf{G}^\ast]_{mn}} \frac{1}{\sigma^2_{\mathsf{c},i}}\textbf{h}_i \textbf{G} \textbf{H}_\mathsf{SR} \textbf{f}_{\mathsf{c}}  +   \textbf{f}_{\mathsf{c}}^{\rm H}\textbf{H}_\mathsf{SR}^{\rm H} \textbf{e}_n \textbf{e}_m^{\rm H} \textbf{h}_i^{\rm H} \frac{1}{\sigma^2_{\mathsf{c},i}}\textbf{h}_i \textbf{G} \textbf{H}_\mathsf{SR} \textbf{f}_{\mathsf{c}} .
\end{align}Now we continue to compute
\begin{align}
\nabla_{[\textbf{G}]_{mn}} \sigma^2_{\mathsf{c},i} &= \sum_{k \, {\in} \, \mathcal{K}} \bar{\alpha} \textbf{h}_i \textbf{G} \textbf{H}_\mathsf{SR} \textbf{f}_k \textbf{f}_k^{\rm H}  \textbf{H}_\mathsf{SR}^{\rm H} \textbf{e}_n \textbf{e}_m^{\rm H} \textbf{h}_i^{\rm H} + \sigma^2\textbf{h}_i \textbf{G} \textbf{e}_n \textbf{e}_m^{\rm H} \textbf{h}_i^{\rm H},
\end{align}as $\nabla_{\textbf{X}}(\textbf{X}\textbf{A}\textbf{X}^{\rm H})= \textbf{X}\textbf{A}$\cite{moon_stirling_math}. Then, we have
\begin{align}
\nabla_{[\textbf{G}]_{mn}}\varepsilon_{\mathsf{c},i}^{\mathsf{min}^{-1}} &= - \sum_{k \, {\in} \, \mathcal{K}} \textbf{f}_{\mathsf{c}}^{\rm H}\textbf{H}_\mathsf{SR}^{\rm H} \textbf{G}^{\rm H} \textbf{h}_i^{\rm H} \frac{1}{\sigma^2_{\mathsf{c},i}}  \bar{\alpha} \textbf{h}_i \textbf{G} \textbf{H}_\mathsf{SR} \textbf{f}_k \textbf{f}_k^{\rm H}  \textbf{H}_\mathsf{SR}^{\rm H} \textbf{e}_n \textbf{e}_m^{\rm H} \textbf{h}_i^{\rm H} \frac{1}{\sigma^2_{\mathsf{c},i}}\textbf{h}_i \textbf{G} \textbf{H}_\mathsf{SR} \textbf{f}_{\mathsf{c}} \nonumber \\
& \quad - \textbf{f}_{\mathsf{c}}^{\rm H}\textbf{H}_\mathsf{SR}^{\rm H} \left( \textbf{G}^{\rm H} \textbf{h}_i^{\rm H} \frac{1}{\sigma^2_{\mathsf{c},i}}\sigma^2\textbf{h}_i \textbf{G} \textbf{e}_n \textbf{e}_m^{\rm H} \textbf{h}_i^{\rm H}\frac{1}{\sigma^2_{\mathsf{c},i}}\textbf{h}_i \textbf{G}  - \textbf{e}_n \textbf{e}_m^{\rm H} \textbf{h}_i^{\rm H} \frac{1}{\sigma^2_{\mathsf{c},i}}\textbf{h}_i \textbf{G} \right)\textbf{H}_\mathsf{SR} \textbf{f}_{\mathsf{c}}, \label{append_invEck_Gc}\\
\nabla_{[\textbf{G}]_{mn}}\mathsf{R}_{\mathsf{c},i} &= - \sum_{k \, {\in} \, \mathcal{K}} \textbf{e}_m^{\rm H} \bar{\alpha} \frac{\varepsilon_{\mathsf{c},i}^{\mathsf{min}}}{\sigma^4_{\mathsf{c},i}} \textbf{h}_i^{\rm H} \textbf{h}_i \textbf{G} \textbf{H}_\mathsf{SR} \textbf{f}_{\mathsf{c}}  \textbf{f}_{\mathsf{c}}^{\rm H}\textbf{H}_\mathsf{SR}^{\rm H} \textbf{G}^{\rm H} \textbf{h}_i^{\rm H}  \textbf{h}_i \textbf{G} \textbf{H}_\mathsf{SR} \textbf{f}_k \textbf{f}_k^{\rm H}  \textbf{H}_\mathsf{SR}^{\rm H} \textbf{e}_n \nonumber \\
& \quad -  \textbf{e}_m^{\rm H} \frac{\varepsilon_{\mathsf{c},i}^{\mathsf{min}} }{\sigma^2_{\mathsf{c},i}} \textbf{h}_i^{\rm H} \textbf{h}_i \textbf{G} \textbf{H}_\mathsf{SR} \textbf{f}_{\mathsf{c}} \textbf{f}_{\mathsf{c}}^{\rm H}\textbf{H}_\mathsf{SR}^{\rm H} \left(\frac{ \sigma^2 }{\sigma^2_{\mathsf{c},i}}  \textbf{G}^{\rm H} \textbf{h}_i^{\rm H}\textbf{h}_i \textbf{G}  -  \textbf{I} \right)\textbf{e}_n .
 \label{grad_R_eps_cl_mn_part1_Gc}
\end{align}Finally, we obtain
\begin{align}
\nabla_{\textbf{G}}\mathsf{R}_{\mathsf{c},i} &= - \sum_{k \, {\in} \, \mathcal{K}}  \bar{\alpha} \frac{\varepsilon_{\mathsf{c},i}^{\mathsf{min}}}{\sigma^4_{\mathsf{c},i}} \textbf{h}_i^{\rm H} \textbf{h}_i \textbf{G} \textbf{H}_\mathsf{SR} \textbf{f}_{\mathsf{c}}  \textbf{f}_{\mathsf{c}}^{\rm H}\textbf{H}_\mathsf{SR}^{\rm H} \textbf{G}^{\rm H} \textbf{h}_i^{\rm H}  \textbf{h}_i \textbf{G} \textbf{H}_\mathsf{SR} \textbf{f}_k \textbf{f}_k^{\rm H}  \textbf{H}_\mathsf{SR}^{\rm H}  \nonumber \\
& \quad -  \frac{ \sigma^2 \varepsilon_{\mathsf{c},i}^{\mathsf{min}}}{\sigma^4_{\mathsf{c},i}} \textbf{h}_i^{\rm H}\textbf{h}_i \textbf{G} \textbf{H}_\mathsf{SR} \textbf{f}_{\mathsf{c}}  \textbf{f}_{\mathsf{c}}^{\rm H}\textbf{H}_\mathsf{SR}^{\rm H} \textbf{G}^{\rm H} \textbf{h}_i^{\rm H}\textbf{h}_i \textbf{G}  +  \frac{\varepsilon_{\mathsf{c},i}^{\mathsf{min}} }{\sigma^2_{\mathsf{c},i}} \textbf{h}_i^{\rm H} \textbf{h}_i \textbf{G} \textbf{H}_\mathsf{SR} \textbf{f}_{\mathsf{c}} \textbf{f}_{\mathsf{c}}^{\rm H}\textbf{H}_\mathsf{SR}^{\rm H} .  \label{grad_R_eps_cl_last_Gc}
\end{align}and the gradient of B becomes
\begin{align}
\nabla_{\textbf{G}} \mathrm{B}
& = \sum_{i \in \mathcal{N}} \sum_{k \, {\in} \, \mathcal{K}} \kappa_i \bar{\alpha} \frac{\varepsilon_{\mathsf{c},i}^{\mathsf{min}}}{\sigma^4_{\mathsf{c},i}} \textbf{h}_i^{\rm H} \textbf{h}_i \textbf{G} \textbf{H}_\mathsf{SR} \textbf{f}_{\mathsf{c}}  \textbf{f}_{\mathsf{c}}^{\rm H}\textbf{H}_\mathsf{SR}^{\rm H} \textbf{G}^{\rm H} \textbf{h}_i^{\rm H}  \textbf{h}_i \textbf{G} \textbf{H}_\mathsf{SR} \textbf{f}_k \textbf{f}_k^{\rm H}  \textbf{H}_\mathsf{SR}^{\rm H}  \nonumber \\
& \quad + \sum_{i \in \mathcal{N}} \kappa_i \frac{ \sigma^2 \varepsilon_{\mathsf{c},i}^{\mathsf{min}}}{\sigma^4_{\mathsf{c},i}} \textbf{h}_i^{\rm H}\textbf{h}_i \textbf{G} \textbf{H}_\mathsf{SR} \textbf{f}_{\mathsf{c}}  \textbf{f}_{\mathsf{c}}^{\rm H}\textbf{H}_\mathsf{SR}^{\rm H} \textbf{G}^{\rm H} \textbf{h}_i^{\rm H}\textbf{h}_i \textbf{G}  - \sum_{i \in \mathcal{N}}  \kappa_i \frac{\varepsilon_{\mathsf{c},i}^{\mathsf{min}} }{\sigma^2_{\mathsf{c},i}} \textbf{h}_i^{\rm H} \textbf{h}_i \textbf{G} \textbf{H}_\mathsf{SR} \textbf{f}_{\mathsf{c}} \textbf{f}_{\mathsf{c}}^{\rm H}\textbf{H}_\mathsf{SR}^{\rm H}.
\label{grad_B_last_Gc}
\end{align}
We conclude by computing the gradient of C and D defined in (\ref{append_lagrangian_SR}) as
\begin{align}
\nabla_{\textbf{G}} \mathrm{C}
&= 0,
\label{grad_C_last_Gc}\\
\nabla_{\textbf{G}} \mathrm{D}
&= \lambda_2 \left( \alpha  \textbf{G} \textbf{H}_\mathsf{SR} \textbf{f}_{\mathsf{c}} \textbf{f}_{\mathsf{c}}^{\rm H} \textbf{H}_\mathsf{SR}^{\rm H}  + \sum_{k \, {\in} \, \mathcal{K}} \bar{\alpha}  \textbf{G} \textbf{H}_\mathsf{SR} \textbf{f}_k \textbf{f}_k^{\rm H} \textbf{H}_\mathsf{SR}^{\rm H} +  \sigma^2  \textbf{G} \textbf{I} \right),
\label{grad_D_last_Gc}
\end{align}
We finally combine (\ref{grad_A_last_Gc}), (\ref{grad_B_last_Gc}),  (\ref{grad_C_last_Gc}), and (\ref{grad_D_last_Gc}) to obtain the desired gradient in \eqref{eq:gradient_wsr_G}.
\section{}\label{sec:gradient_wmmse_F}
In this section, we derive $\nabla_{\textbf{f}_k^\mathsf{\,t}}g\left(\textbf{F}^\mathsf{t},\textbf{G}^\mathsf{t},\theta_\mathsf{c} \right)$ for $\mathsf{\,t} = \mathsf{SC}$. For notational convenience, we drop the superscript $\mathsf{\,t}$. The Lagrangian objective function is given by
\begin{align}
g\left(\textbf{F},\textbf{G},\eta \right) &= \eta + \underbrace{\sum_{n \in \mathcal{N}} \bar{\nu}_n^\mathsf{t} \left( \xi_{\mathsf{u},n}^\mathsf{t,min} - \eta \right)}_\text{A} + \underbrace{\sum_{n \in \mathcal{N}} \bar{\kappa}_n^\mathsf{t} \left(\xi_{\mathsf{c},n}^\mathsf{t,min} - \xi_c^\mathsf{th} \right)}_\text{B} + \bar{\lambda}_1^\mathsf{t} J_1^\mathsf{t} + \bar{\lambda}_2^\mathsf{t} J_2^\mathsf{t}. \label{append_lagrangian_MMSE}
\end{align}Note that $\xi_{\mathsf{u},n}^\mathsf{min} = v_n \, \varepsilon_{\mathsf{u},n}^\mathsf{min} \,{-}\, \log \! \left( \bar{\alpha} v_n \right)$ and $\xi_{\mathsf{c},n}^\mathsf{min} = w_n \, \varepsilon_{\mathsf{c},n}^\mathsf{min} \,{-}\, \log \! \left(\alpha w_n  \right)$. First, we calculate the gradient of A. In (\ref{append_lagrangian_MMSE}) we need $\nabla_{\textbf{f}_k} \varepsilon_{\mathsf{u},i}^\mathsf{min}$ for both $\mu(i)=k$, $(i \in \mathcal{G}_k)$ and $\mu(i) \neq k$, $(i \in \mathcal{N} \setminus \mathcal{G}_k) $. Starting with the case $\mu(i)=k$. Using $\nabla_{\textbf{X}} (\textbf{X}^{-1}) = -\textbf{X}^{-1}\nabla (\textbf{X}) \textbf{X}^{-1}$ \cite{matrix_cookbook} and (\ref{grad_cov_E_eps_k_mn}), we have
\begin{align}
\nabla_{[\textbf{f}_k]_m} \varepsilon_{\mathsf{u},i}^\mathsf{min} &= - \varepsilon_{\mathsf{u},i}^\mathsf{min} \left(\nabla_{[\textbf{f}_k]_m}\varepsilon_{\mathsf{u},i}^\mathsf{min^{-1}}\right) \varepsilon_{\mathsf{u},i}^\mathsf{min}= - \varepsilon_{\mathsf{u},i}^\mathsf{min} \textbf{e}_m^{\rm H} \textbf{H}_\mathsf{SR}^{\rm H} \textbf{G}^{\rm H} \textbf{h}_i^{\rm H} \frac{1}{\sigma^2_{\mathsf{u},i}} \textbf{h}_i \textbf{G} \textbf{H}_\mathsf{SR} \textbf{f}_k \varepsilon_{\mathsf{u},i}^\mathsf{min}.
\end{align}Then we have,
\begin{align}
\nabla_{\textbf{f}_k} \varepsilon_{\mathsf{u},i}^\mathsf{min} &= -  \frac{(\varepsilon_{\mathsf{u},i}^\mathsf{min})^2}{\sigma^2_{\mathsf{u},i}} \textbf{H}_\mathsf{SR}^{\rm H} \textbf{G}^{\rm H} \textbf{h}_i^{\rm H}  \textbf{h}_i \textbf{G} \textbf{H}_\mathsf{SR} \textbf{f}_k\label{grad_E_k}
\end{align}
Next, we compute $\nabla_{\textbf{f}_k} \varepsilon_{\mathsf{u},i}^\mathsf{min}$, for $\mu(i) \neq k$ as
\begin{align}
\nabla_{[\textbf{f}_k]_{m}}\varepsilon_{\mathsf{u},i}^\mathsf{min}
&= - \varepsilon_{\mathsf{u},i}^\mathsf{min} \textbf{f}_i^{\rm H} \textbf{H}_\mathsf{SR}^{\rm H} \textbf{G}^{\rm H} \textbf{h}_i^{\rm H} \nabla_{[\textbf{f}_k]_{m}}\left(\frac{1}{\sigma^2_{\mathsf{u},i}}\right) \textbf{h}_i \textbf{G} \textbf{H}_\mathsf{SR} \textbf{f}_i \varepsilon_{\mathsf{u},i}^\mathsf{min}.\label{grad_cov_E_eps_i_2}
\end{align}Using (\ref{grad_pkm_invrn2i}) and (\ref{grad_pkm_rn2i})
have
\begin{align}
 \nabla_{[\textbf{f}_k]_{m}}\varepsilon_{\mathsf{u},i}^\mathsf{min} &= \bar{\alpha}\textbf{e}_m^{\rm H} \textbf{H}_\mathsf{SR}^{\rm H} \textbf{G}^{\rm H} \textbf{h}_i^{\rm H}\frac{1}{\sigma^2_{\mathsf{u},i}} \textbf{h}_i \textbf{G} \textbf{H}_\mathsf{SR} \textbf{f}_i (\varepsilon_{\mathsf{u},i}^\mathsf{min})^2 \textbf{f}_i^{\rm H} \textbf{H}_\mathsf{SR}^{\rm H} \textbf{G}^{\rm H} \textbf{h}_i^{\rm H} \frac{1}{\sigma^2_{\mathsf{u},i}} \textbf{h}_i \textbf{G} \textbf{H}_\mathsf{SR} \textbf{f}_k. \label{grad_cov_E_eps_i_mn}
\end{align}Then,
\begin{align}
 \nabla_{\textbf{f}_k} \varepsilon_{\mathsf{u},i}^\mathsf{min} &= \bar{\alpha} \frac{(\varepsilon_{\mathsf{u},i}^\mathsf{min})^2}{\sigma^4_{\mathsf{u},i}}  \textbf{H}_\mathsf{SR}^{\rm H} \textbf{G}^{\rm H} \textbf{h}_i^{\rm H} \textbf{h}_i \textbf{G} \textbf{H}_\mathsf{SR} \textbf{f}_i  \textbf{f}_i^{\rm H} \textbf{H}_\mathsf{SR}^{\rm H} \textbf{G}^{\rm H} \textbf{h}_i^{\rm H} \textbf{h}_i \textbf{G} \textbf{H}_\mathsf{SR} \textbf{f}_k. \label{grad_cov_E_eps_i_mn_last}
\end{align}Using (\ref{grad_E_k}) and (\ref{grad_cov_E_eps_i_mn_last}), we conclude that the gradient of A can be written as
\begin{align}
\nabla_{\textbf{f}_k} \mathrm{A} &= \sum_{\substack{ i \in \mathcal{N}\setminus \mathcal{G}_k}} \bar{\nu}_i v_i \bar{\alpha} \frac{(\varepsilon_{\mathsf{u},i}^\mathsf{min})^2}{\sigma^4_{\mathsf{u},i}}  \textbf{H}_\mathsf{SR}^{\rm H} \textbf{G}^{\rm H} \textbf{h}_i^{\rm H} \textbf{h}_i \textbf{G} \textbf{H}_\mathsf{SR} \textbf{f}_i  \textbf{f}_i^{\rm H} \textbf{H}_\mathsf{SR}^{\rm H} \textbf{G}^{\rm H} \textbf{h}_i^{\rm H} \textbf{h}_i \textbf{G} \textbf{H}_\mathsf{SR} \textbf{f}_k\nonumber \\
& \quad -  \sum_{ i \in \mathcal{G}_k} \bar{\nu}_i v_i \frac{(\varepsilon_{\mathsf{u},i}^\mathsf{min})^2}{\sigma^2_{\mathsf{u},i}} \textbf{H}_\mathsf{SR}^{\rm H} \textbf{G}^{\rm H} \textbf{h}_i^{\rm H}  \textbf{h}_i \textbf{G} \textbf{H}_\mathsf{SR} \textbf{f}_k  \label{grad_B_last_2}.
\end{align}Secondly we compute the gradient of B defined in (\ref{append_lagrangian_MMSE}). Note that  $\nabla_{\textbf{f}_k} \varepsilon_{\mathsf{c},n}^\mathsf{min} = - \varepsilon_{\mathsf{c},n}^\mathsf{min} (\nabla_{\textbf{f}_k} \varepsilon_{\mathsf{c},n}^\mathsf{min^{-1}}) \varepsilon_{\mathsf{c},n}^\mathsf{min}$. The gradient $\nabla_{[\textbf{f}_k]_{m}} \varepsilon_{\mathsf{c},n}^\mathsf{min^{-1}}$ is calculated in (\ref{append_invEck}). Then we have,
\begin{align}
\nabla_{\textbf{f}_k} \varepsilon_{\mathsf{c},n}^\mathsf{min} &= \bar{\alpha} \frac{(\varepsilon_{\mathsf{c},n}^\mathsf{min} )^2}{\sigma^4_{\mathsf{c},n}} \textbf{H}_\mathsf{SR}^{\rm H} \textbf{G}^{\rm H} \textbf{h}_n^{\rm H} \textbf{h}_n \textbf{G} \textbf{H}_\mathsf{SR} \textbf{f}_{\mathsf{c}}  \textbf{f}_{\mathsf{c}}^{\rm H}\textbf{H}_\mathsf{SR}^{\rm H} \textbf{G}^{\rm H} \textbf{h}_n^{\rm H} \textbf{h}_n \textbf{G} \textbf{H}_\mathsf{SR} \textbf{f}_k \nonumber \\
&\quad - \frac{(\varepsilon_{\mathsf{c},n}^\mathsf{min} )^2}{\sigma^2_{\mathsf{c},n}}  \textbf{H}_\mathsf{SR}^{\rm H} \textbf{G}^{\rm H} \textbf{h}_n^{\rm H} \textbf{h}_n \textbf{G} \textbf{H}_\mathsf{SR} \textbf{f}_{\mathsf{c}}. \label{append_invEck_2}
\end{align}The gradient of B becomes
\begin{align}
\nabla_{\textbf{f}_k} \mathrm{B}
 &= \sum_{n \in \mathcal{N}} \bar{\kappa}_n w_n  \bar{\alpha} \frac{(\varepsilon_{\mathsf{c},n}^\mathsf{min} )^2}{\sigma^4_{\mathsf{c},n}} \textbf{H}_\mathsf{SR}^{\rm H} \textbf{G}^{\rm H} \textbf{h}_n^{\rm H} \textbf{h}_n \textbf{G} \textbf{H}_\mathsf{SR} \textbf{f}_{\mathsf{c}}  \textbf{f}_{\mathsf{c}}^{\rm H}\textbf{H}_\mathsf{SR}^{\rm H} \textbf{G}^{\rm H} \textbf{h}_n^{\rm H} \textbf{h}_n \textbf{G} \textbf{H}_\mathsf{SR} \textbf{f}_k  \nonumber \\
&\quad - \sum_{n \in \mathcal{N}} \bar{\kappa}_n w_n  \frac{(\varepsilon_{\mathsf{c},n}^\mathsf{min} )^2}{\sigma^2_{\mathsf{c},n}} \textbf{H}_\mathsf{SR}^{\rm H} \textbf{G}^{\rm H} \textbf{h}_n^{\rm H} \textbf{h}_n \textbf{G} \textbf{H}_\mathsf{SR} \textbf{f}_{\mathsf{c}}  .
\label{grad_A_last_2}
\end{align}
 Finally combining (\ref{grad_C_last}),  (\ref{grad_D_last}), (\ref{grad_B_last_2}) and (\ref{grad_A_last_2}), we 
 readily obtain \eqref{grad_g_append}.

\section{}\label{sec:gradient_wmmse_G}
In this section, we derive $\nabla_{\textbf{G}^\mathsf{t}}g\left(\textbf{F}^\mathsf{t},\textbf{G}^\mathsf{t},\eta \right)$ for $\mathsf{\,t} = \mathsf{SC}$. For notational convenience, we drop the superscript $\mathsf{\,t}$. Note that $\xi_{\mathsf{u},n}^\mathsf{min} = v_n \, \varepsilon_{\mathsf{u},n}^\mathsf{min} \,{-}\, \log \! \left( \bar{\alpha} v_n \right)$ and $\xi_{\mathsf{c},n}^\mathsf{min} = w_n \, \varepsilon_{\mathsf{c},n}^\mathsf{min} \,{-}\, \log \! \left(\alpha w_n  \right)$.
First, we compute the gradient of A in (\ref{append_lagrangian_MMSE}). We have $\nabla_{\textbf{G}} \varepsilon_{\mathsf{u},i}^\mathsf{min} = -(\varepsilon_{\mathsf{u},i}^\mathsf{min})^2 \nabla_{\textbf{G}}\varepsilon_{\mathsf{u},i}^{\mathsf{min}^{-1}} $. Note that the noise variance $\frac{1}{\sigma^2_{\mathsf{u},i}}$ is not independent from $\textbf{G}$. Thus, the gradient $\nabla_{[\textbf{G}]_{mn}} \varepsilon_{\mathsf{u},i}^{\mathsf{min}^{-1}}$ is calculated in (\ref{append_invEck_G}). Then, we have
\begin{align}
\nabla_{\textbf{G}}\varepsilon_{\mathsf{u},i}^\mathsf{min} &= \sum_{\substack{k \, {\in} \, \mathcal{K}\\ k\neq \mu(i)}}  \bar{\alpha}  \frac{(\varepsilon_{\mathsf{u},i}^\mathsf{min})^2}{\sigma^4_{\mathsf{u},i}} \textbf{h}_i^{\rm H} \textbf{h}_i \textbf{G} \textbf{H}_\mathsf{SR} \textbf{f}_{\mu(i)}  \textbf{f}_{\mu(i)}^{\rm H}\textbf{H}_\mathsf{SR}^{\rm H} \textbf{G}^{\rm H} \textbf{h}_i^{\rm H} \textbf{h}_i \textbf{G} \textbf{H}_\mathsf{SR} \textbf{f}_k \textbf{f}_k^{\rm H}  \textbf{H}_\mathsf{SR}^{\rm H} \nonumber \\
& \quad + \sigma^2 \frac{(\varepsilon_{\mathsf{u},i}^\mathsf{min})^2}{\sigma^4_{\mathsf{u},i}} \textbf{h}_i^{\rm H}\textbf{h}_i \textbf{G} \textbf{H}_\mathsf{SR} \textbf{f}_{\mu(i)} \textbf{f}_{\mu(i)}^{\rm H}\textbf{H}_\mathsf{SR}^{\rm H} \textbf{G}^{\rm H} \textbf{h}_i^{\rm H} \textbf{h}_i \textbf{G}  -  \frac{(\varepsilon_{\mathsf{u},i}^\mathsf{min})^2}{\sigma^2_{\mathsf{u},i}}\textbf{h}_i^{\rm H}\textbf{h}_i \textbf{G} \textbf{H}_\mathsf{SR} \textbf{f}_{\mu(i)} \textbf{f}_{\mu(i)}^{\rm H}\textbf{H}_\mathsf{SR}^{\rm H} \label{append_invEck_G_2}
\end{align}and the gradient of A becomes
\begin{align}
\nabla_{\textbf{G}} \mathrm{A}
& = \sum_{i \in \mathcal{N}} \sum_{\substack{k \, {\in} \, \mathcal{K}\\ k\neq \mu(i)}}  \bar{\nu}_i v_i \bar{\alpha}  \frac{(\varepsilon_{\mathsf{u},i}^\mathsf{min})^2}{\sigma^4_{\mathsf{u},i}} \textbf{h}_i^{\rm H} \textbf{h}_i \textbf{G} \textbf{H}_\mathsf{SR} \textbf{f}_{\mu(i)}  \textbf{f}_{\mu(i)}^{\rm H}\textbf{H}_\mathsf{SR}^{\rm H} \textbf{G}^{\rm H} \textbf{h}_i^{\rm H} \textbf{h}_i \textbf{G} \textbf{H}_\mathsf{SR} \textbf{f}_k \textbf{f}_k^{\rm H}  \textbf{H}_\mathsf{SR}^{\rm H} \nonumber \\
& \quad + \sum_{i \in \mathcal{N}} \bar{\nu}_i v_i \frac{(\varepsilon_{\mathsf{u},i}^\mathsf{min})^2}{\sigma^2_{\mathsf{u},i}}\textbf{h}_i^{\rm H}\textbf{h}_i \textbf{G} \textbf{H}_\mathsf{SR} \textbf{f}_{\mu(i)} \textbf{f}_{\mu(i)}^{\rm H}\textbf{H}_\mathsf{SR}^{\rm H} \left(\frac{\sigma^2}{\sigma^2_{\mathsf{u},i}}  \textbf{G}^{\rm H} \textbf{h}_i^{\rm H} \textbf{h}_i \textbf{G}  -  \textbf{I} \right).
\label{grad_A_last_Gc_2}
\end{align}
The next one is to compute the gradient of B and we have $\nabla_{\textbf{G}} \varepsilon_{\mathsf{c},i}^\mathsf{min} = - \varepsilon_{\mathsf{c},i}^\mathsf{min} \nabla_{\textbf{G}}\varepsilon_{\mathsf{c},i}^{\mathsf{min}^{-1}} \varepsilon_{\mathsf{c},i}^\mathsf{min}$. Note that the noise variance $\frac{1}{\sigma^2_{\mathsf{c},i}}$ is not independent from $\textbf{G}$. Thus, the gradient $\nabla_{[\textbf{G}]_{mn}}\varepsilon_{\mathsf{c},i}^{\mathsf{min}^{-1}}$ is calculated in (\ref{append_invEck_Gc}). Then, we have
\begin{align}
\nabla_{\textbf{G}}\varepsilon_{\mathsf{c},i}^\mathsf{min} &=  \sum_{k \, {\in} \, \mathcal{K}} \bar{\alpha} \frac{(\varepsilon_{\mathsf{c},i}^\mathsf{min})^2}{\sigma^4_{\mathsf{c},i}}\textbf{h}_i^{\rm H}\textbf{h}_i \textbf{G} \textbf{H}_\mathsf{SR} \textbf{f}_{\mathsf{c}}\textbf{f}_{\mathsf{c}}^{\rm H}\textbf{H}_\mathsf{SR}^{\rm H} \textbf{G}^{\rm H} \textbf{h}_i^{\rm H} \textbf{h}_i \textbf{G} \textbf{H}_\mathsf{SR} \textbf{f}_k \textbf{f}_k^{\rm H}  \textbf{H}_\mathsf{SR}^{\rm H}  \nonumber \\
& \quad +  \sigma^2 \frac{(\varepsilon_{\mathsf{c},i}^\mathsf{min})^2}{\sigma^4_{\mathsf{c},i}}\textbf{h}_i^{\rm H}\textbf{h}_i \textbf{G} \textbf{H}_\mathsf{SR} \textbf{f}_{\mathsf{c}} \textbf{f}_{\mathsf{c}}^{\rm H}\textbf{H}_\mathsf{SR}^{\rm H} \textbf{G}^{\rm H} \textbf{h}_i^{\rm H} \textbf{h}_i \textbf{G}  -  \frac{(\varepsilon_{\mathsf{c},i}^\mathsf{min})^2}{\sigma^2_{\mathsf{c},i}}\textbf{h}_i^{\rm H}\textbf{h}_i \textbf{G} \textbf{H}_\mathsf{SR} \textbf{f}_{\mathsf{c}}\textbf{f}_{\mathsf{c}}^{\rm H}\textbf{H}_\mathsf{SR}^{\rm H}. \label{append_invEck_Gc_2}
\end{align}The gradient of B becomes
\begin{align}
\nabla_{\textbf{G}} \mathrm{B}
& = \sum_{i \in \mathcal{N}} \sum_{k \, {\in} \, \mathcal{K}}\bar{\kappa}_i w_i \bar{\alpha} \frac{(\varepsilon_{\mathsf{c},i}^\mathsf{min})^2}{\sigma^4_{\mathsf{c},i}}\textbf{h}_i^{\rm H}\textbf{h}_i \textbf{G} \textbf{H}_\mathsf{SR} \textbf{f}_{\mathsf{c}}\textbf{f}_{\mathsf{c}}^{\rm H}\textbf{H}_\mathsf{SR}^{\rm H} \textbf{G}^{\rm H} \textbf{h}_i^{\rm H} \textbf{h}_i \textbf{G} \textbf{H}_\mathsf{SR} \textbf{f}_k \textbf{f}_k^{\rm H}  \textbf{H}_\mathsf{SR}^{\rm H}  \nonumber \\
& \quad + \sum_{i \in \mathcal{N}} \bar{\kappa}_i w_i \frac{(\varepsilon_{\mathsf{c},i}^\mathsf{min})^2}{\sigma^2_{\mathsf{c},i}}\textbf{h}_i^{\rm H}\textbf{h}_i \textbf{G} \textbf{H}_\mathsf{SR} \textbf{f}_{\mathsf{c}}\textbf{f}_{\mathsf{c}}^{\rm H}\textbf{H}_\mathsf{SR}^{\rm H}\left( \frac{\sigma^2}{\sigma^2_{\mathsf{c},i}} \textbf{G}^{\rm H} \textbf{h}_i^{\rm H} \textbf{h}_i \textbf{G}  - \textbf{I} \right).
\label{grad_B_last_Gc_2}
\end{align}
Finally combining (\ref{grad_C_last_Gc}),  (\ref{grad_D_last_Gc}), (\ref{grad_A_last_Gc_2}) and (\ref{grad_B_last_Gc_2}) we readily obtain \eqref{grad_f_append_Gc_2}.

\end{appendices}

\bibliographystyle{IEEEtran} 
\bibliography{references}
\end{document}